\documentclass[lettersize,journal]{IEEEtran}
\usepackage{amsmath,amsfonts}
\usepackage{algorithmic}
\usepackage{algorithm}
\usepackage{array}
\usepackage[caption=false,font=normalsize,labelfont=sf,textfont=sf]{subfig}
\usepackage{textcomp}
\usepackage{stfloats}
\usepackage{url}
\usepackage{verbatim}
\usepackage{graphicx}
\usepackage{cite}
\usepackage{booktabs}
\usepackage{float}
\usepackage{multirow}
\usepackage{xcolor}
\usepackage[normalem]{ulem}
\usepackage{arydshln}

\DeclareMathOperator{\EX}{\mathbb{E}}

\begin{document}

\title{An Efficient End-to-End Approach to Noise Invariant Speech Features via Multi-Task Learning}

\author{
Heitor~R.~Guimarães,~\IEEEmembership{Student Member,~IEEE,}
Arthur~Pimentel,~\IEEEmembership{}
Anderson~R.~Avila,~\IEEEmembership{Member,~IEEE,}
Mehdi~Rezagholizadeh,~\IEEEmembership{}
Boxing~Chen,~\IEEEmembership{}
and Tiago~H.~Falk,~\IEEEmembership{Senior Member,~IEEE}}



\maketitle

\begin{abstract}
Self-supervised speech representation learning enables the extraction of meaningful features from raw waveforms. These features can then be efficiently used across multiple downstream tasks. However, two significant issues arise when considering the deployment of such methods ``in-the-wild": (i) Their large size, which can be prohibitive for edge applications; and (ii) their robustness to detrimental factors, such as noise and/or reverberation, that can heavily degrade the performance of such systems. In this work, we propose RobustDistiller, a novel knowledge distillation mechanism that tackles both problems jointly. Simultaneously to the distillation recipe, we apply a multi-task learning objective to encourage the network to learn noise-invariant representations by denoising the input. The proposed mechanism is evaluated on twelve different downstream tasks. It outperforms several benchmarks regardless of noise type, or noise and reverberation levels. Experimental results show that the new Student model with 23M parameters can achieve results comparable to the Teacher model with 95M parameters. Lastly, we show that the proposed recipe can be applied to other distillation methodologies, such as the recent DPWavLM. For reproducibility, code and model checkpoints will be made available at \mbox{\url{https://github.com/Hguimaraes/robustdistiller}}.

\end{abstract}


\section{Introduction}
\IEEEPARstart{S}{elf}-supervised speech representation learning (S3RL) has become commonplace in speech processing applications. These methods allow us to learn meaningful and disentangled universal features from high-dimensional data. These features are attained in a self-supervised manner by exploiting the inner structure of the data modality through a pretext task~\cite{9893562}. Later, these universal representations serve as input to different downstream systems, such as speech and/or speaker recognizers. Today, \mbox{Wav2Vec 2.0}~\cite{baevski2020wav2vec}, HuBERT~\cite{hsu2021hubert}, and WavLM~\cite{chen2022wavlm} are the most widely used models to extract cross-task speech representations.

Universal representations can be important for edge applications, where storage capacity may be limited and storing the weights of different models is unfeasible. Universal speech representations, however, face certain limitations in edge applications. First, their performance drastically drops in unseen environmental (test) conditions not seen during training \cite{huang2022improving}. Second, in some configurations, the representations rely on a very large number of parameters, thus making them impractical for resource-constrained devices. For instance, while the HuBERT-\emph{base}~\cite{hsu2021hubert} model has 95 million parameters, the XLS-R~\cite{babu2021xls}, a multilingual model, can range up to 2 billion.

To overcome these limitations, different techniques have been applied. For example, the work in \cite{huang2022improving} showed that the HuBERT representation can be more robust to unseen conditions via domain adversarial training and data augmentation. Therefore, the so-called ``Robust HuBERT" model was developed. Alternately, different compression techniques have been explored to reduce the footprint of those models, with promising results seen using knowledge distillation (KD). For example, the work in \cite{chang2022distilhubert} showed that HuBERT could be compressed by almost a factor of four via KD, resulting in the so-called DistilHuBERT model. Notwithstanding, recent results have shown that DistilHuBERT has sensitivity to unseen test conditions \cite{guimarães2023exploration}, especially those related to environmental noise and reverberation. New compression techniques using distillation and pruning have been shown to improve performance in clean settings. For instance, the DPWavLM model~\cite{peng23c_interspeech} benefits from these innovations. Furthermore, experiments show these methods can reduce gaps between compressed and larger models in noise. However, there is still room for improvement, especially on speaker-related downstream tasks. As such, to date, only a handful of systems have successfully combined model compression and environmental robustness for edge speech applications. This paper aims to fill this gap.

More specifically, the present study expands on the authors' previous work \cite{robustdistiller} and proposes RobustDistiller. The novel training recipe was designed to compress universal models while making them more resilient to unseen noise perturbations commonly encountered ``in the wild''. Our new contributions can be summarized as follows:

\begin{enumerate}
    \item We propose a feature denoising KD recipe that helps the Student model learn noise-invariant features, making speech applications more robust in noisy environments.
    \item Through extensive experimentation, we show that the denoising step introduced during distillation gives the Student model better disentanglement properties. Hence, this modification leads to better generalization.
    \item The proposed recipe is tested across representations (i.e., HuBERT, Wav2Vec 2.0, WavLM, and Robust HuBERT~\cite{huang2022improving}), showing its generalizability.
    \item Empirically, our recipe shows that a Student model can outperform its robust Teacher in noisy scenarios for content-related tasks with SNR ranging from $[-5, 5]$ dB.
    \item Lastly, although our method is built on top of the DistilHuBERT for most experiments, the proposed recipe is shown to be flexible and to work alongside other compression strategies, such as \mbox{DPWavLM}~\cite{peng23c_interspeech}.
\end{enumerate}

The remainder of this paper is organized as follows. Section II presents background material and related works. Section III describes the proposed method, while Section IV presents the experimental setup. Section V presents the obtained results and discusses them in light of existing literature. Finally, Section VII presents the conclusions and future research directions.

\section{Background and Related Work}

\subsection{Self-Supervised Speech Representation Learning}
Learning to extract relevant features in an unsupervised manner is a challenging machine-learning task. Recently, the Wav2Vec 2.0~\cite{baevski2020wav2vec} was proposed as a specific model for S3RL. The key principle behind the model is to learn acoustic features directly from speech utterances. The architecture consists of a multi-layer convolutional encoder that outputs intermediate latent speech representations for every 20 ms of audio. Next, some latent representations are randomly masked, as in a masked language modeling (MLM) problem. The context network, i.e., a large Transformer encoder, then attempts to predict the discrete speech units from a codebook obtained through a Gumbel-Softmax distribution \cite{jang2017categorical}. The Wav2Vec 2.0 model is optimized via a contrastive loss.

Predictive pretext tasks are another popular paradigm for learning representations from speech signals. In this category, the model directly relies on a multi-class classification approach. Today, HuBERT and WavLM are representative state-of-the-art models in this category, showing top performances across several tasks. The HuBERT and WavLM architectures are very similar to Wav2Vec 2.0 and consist of a convolutional feature encoder and a large Transformer encoder.

HuBERT's training objective is to solve a masked-language modeling (MLM) task. But instead of directly predicting the continuous representation from the masked acoustic frame, the model supervisory signal is a discrete phonetic unit obtained via an offline K-means clustering. Finally, a BERT-like prediction loss is applied over the masked regions, forcing the model to learn a combined acoustic and language model. 

Similarly to HuBERT, WavLM~\cite{chen2022wavlm} is also a predictive model, but shown to be more robust against environmental factors. This was achieved by training the model on a large and diverse dataset, not constrained to only  audiobooks, as other predecessors. Moreover, two augmentation strategies are performed during the model training phase: (i) add overlapping speech to the utterances; and (ii) contaminate the signals with additive noise. Thus, in addition to the predictive task, the model needs to implicitly perform a speech-denoising step to be able to predict the discrete units correctly.

\subsection{Speech processing Universal PERformance Benchmark}

As more universal representations emerge, the need for a standard method to compare them is needed. The Speech processing Universal PERformance Benchmark (SUPERB)~\cite{yang2021superb} is a leaderboard to benchmark the performance of new universal representations on a set of 10 tasks in a standardized manner. The ten tasks fall under four main categories: content, speaker, semantic, and paralinguistic. Under the content category, tasks include keyword spotting (\textbf{KS}), phoneme recognition (\textbf{PR}), automatic speech recognition (\textbf{ASR}), and query-by-example (\textbf{QbE}). The speaker category includes speaker identification (\textbf{SID}), automatic speaker verification (\textbf{ASV}), and speaker diarization (\textbf{SD}). Semantic tasks, in turn, include spoken intent classification (\textbf{IC}) and spoken slot filling (\textbf{SF}). Finally, the paralinguistic task include speech emotion recognition (\textbf{ER}). The interested reader is referred to~\cite{yang2021superb} for more details about the SUPERB tasks and the datasets used.

More recently, the SUPERB-Speech Generation (SG) benchmark has been proposed in~\cite{tsai-etal-2022-superb}. It extends the tasks from the original SUPERB to include generative tasks. This new benchmark includes speech translation, speech enhancement, source separation, voice conversion, and out-of-distribution ASR. We are particularly interested in two downstream tasks relevant to ``in-the-wild'' scenarios, i.e., speech enhancement (\textbf{SE}) and source separation (\textbf{SS}).

Here, we also adopt the ``score'' metric used in~\cite{feng2023superb} to compute a general score across all the different tasks of the SUPERB challenge, thus allowing the assessment of a  representation's success across tasks. This is important, as different tasks can have different figures-of-merit (e.g., accuracy or equal error rate), thus simple averaging or summing would not suffice. The score metric is given by:
\begin{equation}
    \mathrm{Score}(u) = \frac{1000}{|T|}\Sigma_t^T \frac{1}{|I_t|}\Sigma_i^{I_t} \frac{s_{t,i}(u) - s_{t,i}(\mathrm{base})}{s_{t,i}(\mathrm{SOTA}) - s_{t,i}(\mathrm{base})},
\end{equation}
where $u$ is the upstream model being evaluated (e.g., WavLM), $s_{t,i}$ is the $i$-th metric from the task $t$. Following this nomenclature, $s_{t,i}(\mathrm{SOTA})$ and $s_{t,i}(\mathrm{base})$ represent the best and worst metrics achieved in the given task $t$ for the metric $i$, respectively. $|T|$ is the total of tasks being measured and $|I_t|$ the total of metrics reported for that given task. The final score ranges from 0 to 1000, with higher values indicating better overall model performance across all tasks.

\subsection{Model Compression and Noise Robustness: Prior Art}

While S3RL has become ubiquitous across speech tasks, models are still sensitive to conditions typically observed in edge applications. For example, recent research has shown that when test data is corrupted by factors not seen during training (e.g., reverberation and noise), the distribution shift hampers model performance \cite{robustdistiller}. This has been shown to be common across speech representations, as they are usually trained on extensive unlabelled (clean) data from audiobook readings \cite{panayotov2015librispeech, kahn2020libri}. Techniques that have been explored to mitigate this out-of-domain issue include domain adaptation and domain generalization \cite{hsu2021robust}. Robust HuBERT~\cite{huang2022improving}, for example, proposes to use domain adversarial training (DAT)~\cite{JMLR:v17:15-239} to learn noise-invariant features, using the HuBERT model as a baseline, and improved robustness to noisy test conditions have been reported \cite{huang2022improving}.

Moreover, edge applications are commonly limited in storage capacity and processing power. For example, the HuBERT model can range from 95 million to 1 billion 32-bit float parameters \cite{hsu2021hubert} due to the size of its Transformer network. Such requirements may place strain on edge devices. To counter this issue, different compression schemes have been explored. DistilHuBERT \cite{chang2022distilhubert}, for example, relies on knowledge distillation to compress the HuBERT model by almost a factor of four. More specifically, the layerwise distillation mechanism occurs with a Student model of only two Transformer layers. The representation is fed to three prediction heads that approximate the latent representations from the \{4\textsuperscript{th}, 8\textsuperscript{th}, and 12\textsuperscript{th}\} layers of the Teacher model. After training, the prediction heads are discarded. DistilHuBERT optimizes a loss function that combines reconstruction elements with a cosine similarity metric. 
LightHuBERT~\cite{wang2022lighthubert}, in turn, is an alternative compression methodology for HuBERT which consists of pruning structured parameters and a two-step distillation mechanism. Although LightHuBERT is currently one of the best compression methodologies for S3RL models, their recipe is computationally expensive and can be prohibitive to train on academic resources.

Due to its ease of use, DistilHuBERT has been extended in recent works to improve its performance under clean settings. Two similar approaches, FitHuBERT~\cite{lee22p_interspeech} and the \textit{Deep-versus-Wide} (DW)~\cite{ashihara22_interspeech} method, were developed concomitantly as an extension of the DistilHuBERT recipe. In their work, the authors propose deeper Transformer encoder networks for the Student model while making it thinner. Furthermore, DW further explores the use of layer-to-layer (L2L) prediction to compute the similarity between Teacher and Student embeddings, unlike  DistilHuBERT which uses prediction heads only on top of the last Transformer layer. Empirically, the authors show that L2L distillation for deeper models can work better than the original DistilHuBERT.

Lastly, a more recent work termed DPHuBERT~\cite{peng23c_interspeech} was proposed using distillation and structured pruning techniques (DP) to achieve better model performance on several tasks from SUPERB. The training consists of a two-step recipe. First, a Student model is created as a copy of the Teacher model. In this step, during the distillation, the Student weights are regularized such that 75\% of them are eliminated with the structured pruning method. In the second step, a final distillation procedure is performed to keep the representations from the small Student model as similar to the Teacher's as possible.

As can be seen, the methods discussed above either focus on improving the robustness of universal representations to unseen conditions or on enabling the compression of model sizes. To the best of our knowledge, only one work~\cite{10022474}, developed concurrently with ours, has investigated the compression and robustness of S3RL models. Inspired by the Robust HuBERT model, the authors combine domain adversarial training (DAT) and the bootstraping-your-own-latents (BYOL)~\cite{grill2020bootstrap} style of training. Specifically, the Teacher and Student networks receive different distorted views of the same utterance as input. Unlike our work, the HuBERT model is distilled and evaluated only on three downstream tasks. We also encountered three key differences from our work: (i) We learn noise-invariance via a multi-task learning approach instead of DAT, and only the Student model receives distorted inputs; (ii) we show that the RobustDistiller can be applied to other Teacher models beyond just HuBERT; and (iii) we extend our analysis to all ten tasks from the SUPERB benchmark and two tasks from the SUPERB-SG benchmark. Our approach, termed RobustDistiller, extends both the DistilHuBERT and DPHuBERT recipes to generate robust universal representations. The proposed method is described in detail next.

\begin{figure*}
        \centering
        \includegraphics[width=\linewidth]{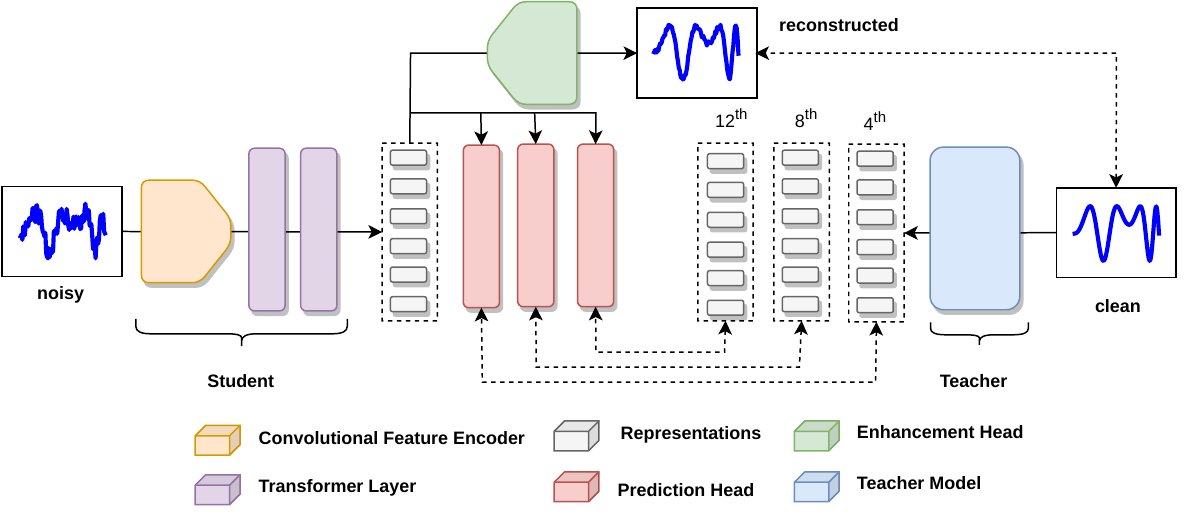} 
        \caption{Diagram of the proposed RobustDistiller recipe. The input of the Student model can be a noisy waveform. The Teacher, on the other hand, receives the clean waveform. The Student must learn clean features and simultaneously retrieve as much information as possible to reconstruct a clean utterance.}
        \label{fig:robust_distiller_diagram}
\end{figure*}

\section{RobustDistiller: Joint distillation and noise-invariant representation extraction }

The RobustDistiller recipe was first introduced in~\cite{robustdistiller} and shown to compress the HuBERT-\emph{base} representation to a quarter of its original size, thus reducing the number of parameters from 94 million to 23 million approximately. This was achieved while creating a more robust model against environmental artifacts. Initially, RobustDistiller was proposed as an extension of the DistilHuBERT recipe with a robustness-first principle. Herein, we show that our proposed recipe can be extended to other distillation recipes, as discussed and empirically shown in the following sections. Two main modifications are proposed to improve the robustness of the Student models: (i) a feature denoising knowledge distillation step that induces the Student model to learn noise-invariant representations; and (ii) a multi-task learning approach via a signal enhancement step where, given the last hidden state from the Student model obtained from a noisy signal, we reconstruct the clean waveform or estimate a binary mask to be applied on the STFT of the noisy input. Figure~\ref{fig:robust_distiller_diagram} shows the schematics of the proposed model. More details about these modifications are given next.

\subsection{Modification \#1: Feature Denoising Knowledge Distillation}

As stated earlier, one of the main challenges that S3RL models face is domain shift, where the training and real-world test data differ significantly in e.g., channel conditions and environmental noises. This can result in system performance degradation, rendering the models less effective for practical use. Data augmentation has been proposed to address this issue by increasing data diversity and reducing model bias \cite{6639100}. 

Inspired by the WavLM training scheme, here we introduce an online contamination method during the distillation process. Our method involves training a Student model to reconstruct the clean representations of a Teacher model while being fed with noisy data as input. During training, a batch of clean speech utterances is uniformly sampled and subjected to one of the following actions: (i) no changes are made to the training utterance; (ii) the utterance is contaminated with additive noise with a signal-to-noise ratio (SNR) chosen randomly from $[0, 20]$ dB; (iii) the speech waveform is convolved with a random room impulse response; and (iv) both noise and reverberation are applied simultaneously. Mathematically, the goal of feature denoising knowledge distillation is to minimize the distance between the Student and Teacher's representation, as follows:
\begin{equation}
    \min_\theta~\EX_{x\sim\mathcal{D}}\left[\frac{1}{m}\sum_{i=1}^m\sum_{\ell \in L} \mathcal{L}_D\left(f_T^{(\ell)}(\pmb{x}_i; \phi), f_S^{(\ell)}(\hat{\pmb{x}}_i; \theta)\right) \right],
\end{equation}
where $\pmb{x}_i$ is sample from an unlabeled dataset $\mathcal{D} = \{x_i \mid i = 1, ..., m\}$ and $\hat{\pmb{x}}_i$ is the distorted view of such sample. $f_T^{(\ell)}$ is the hidden representation of the $\ell$-th layer from the Teacher model parametrized by $\phi$, and $f_S^{(\ell)}$ is the Student prediction that approximates the Teacher representation. $L$ is the set of layers chosen to be distilled (e.g., 4\textsuperscript{th}, 8\textsuperscript{th}, and 12\textsuperscript{th} if following the DistilHuBERT recipe). Lastly, $\mathcal{L}_D$ follows the same definition of~\cite{chang2022distilhubert} as a combination of L1 loss and a cosine similarity between representations.

With this feature denoising step, our central hypothesis is that the model must learn to disentangle the speech signal information from the background noise, leveraging more robust (noise-invariant) features. Note that this methodology can be applied to both layer-wise distillation and L2L methods.

\subsection{Modification \#2: Multi-task learning}
Speech enhancement is a fundamental task in speech processing aimed at improving the quality of a speech signal by removing detrimental factors. This paper proposes a multi-task learning approach inspired by speech enhancement and motivated by recent advancements in speech representation learning \cite{wang2022improving}. In this setting, our proposed approach involves reconstructing the clean waveform from the learned representation of the Student model via an additional enhancement head in complement to the layerwise distillation process. Unlike traditional enhancement techniques, we aim to ensure that the upstream model carries sufficient information about the speech signal rather than focusing on optimizing a quality metric.

To achieve this goal, we explored two architectures for the enhancement head. First, we experiment with an enhancement head similar to the encoder of the Student model. Specifically, we employ a BiLSTM layer followed by seven transposed convolutions and GELU activation functions. This method aims to reconstruct the waveform directly from the last hidden state of our Student model. The main limitation of this proposed head is that the enhancement head is large, with roughly 17 million parameters. Although we will discard the enhancement head after training, it can be a memory bottleneck during training.

Next, we also experiment with a lightweight enhancement head that operates in the time-frequency domain. This enhancement head has only 5 million parameters, not impacting the original distillation time too much. Herein, we estimate a binary mask that will be applied to the magnitude of the noisy signal's short-time Fourier transform (STFT). Later, we reconstruct the clean waveform to measure the quality of the enhanced signal (e.g., PESQ~\cite{941023}) by using the noisy phase of the signal. This enhancement head consists of a single BiLSTM network with three layers and a sigmoid activation function to create the binary mask.

Additionally, we experiment with three different loss functions: L1-, L2-, and a multi-resolution STFT-loss to guide the enhancement head. In our experiments, we will discuss the impact of each choice in more detail. We refer to the combination of these two procedures, i.e., feature denoising knowledge distillation and multi-task learning, as the RobustDistiller recipe, which can be applied across unsupervised representations. 
In our experiments, we build on the work of ~\cite{robustdistiller} and apply the RobustDistiller recipe to distill four popular speech representations: \emph{Wav2Vec 2.0}~\cite{baevski2020wav2vec}, HuBERT \cite{hsu2021hubert}, WavLM \cite{chen2022wavlm}, as well as the more recent Robust HuBERT\cite{huang2022improving}.

\subsection{Notes on extending RobustDistiller to DP-methods}

It is interesting to notice that, although RobustDistiller's recipe was originally developed on top of the DistilHuBERT method, we can expand this idea to other methods as well. To test this hypothesis, in this work we explore expanding the DPWavLM model with our modifications. More specifically, our proposed modifications are inserted during the second distillation step of the recipe, after the distillation-plus-pruning step. We opted for this implementation to reduce the computational cost; however, it is an interesting future work to explore the effects of robust distillation on both steps.

\section{Experimental Setup}
\subsection{Datasets} \label{sec:data}
Similar to DistilHuBERT, the distillation recipe of the proposed model is based on the clean speech utterances of the LibriSpeech corpus \cite{panayotov2015librispeech}, a dataset with 960 hours of audiobook recordings with a 16 kHz sampling rate derived from the LibriVox project. Additionally, to augment the training data, two noise datasets are used, namely MUSAN \cite{musan2015} and UrbanSound8K \cite{Salamon:UrbanSound:ACMMM:14}, as well as the OpenSLR28 dataset \cite{ko2017study} of room impulse responses (RIR). The noise datasets contain approximately 15 hours of recordings in a wide variety of categories, from office-like noises, natural sounds, babble noise, and urban sounds, such as engine idling and siren. We removed the children playing and street music categories to focus on non-speech like noise sources in this first analysis. All the utterances are resampled to 16 kHz. Lastly, the OpenSLR28 dataset contains 325 RIRs recorded in real environments, from small meeting rooms to large auditoriums.

At test time, two additional noise datasets are used to test the model performance under unseen conditions. The first is the Acoustic Scene Classification from the Detection and Classification of Acoustic Scenes and Events Challenge (DCASE2020) \cite{Mesaros2018_DCASE}, comprised of 64 hours of audio recordings in 10 acoustic scenes, with four different recording devices in 12 cities. For the reverberation condition, we used the OpenSLR26 dataset \cite{ko2017study}, which contains 60,000 simulated room impulse responses corresponding to various small-, medium-, and large-sized rooms.

\subsection{Pre-training}
Here, we evaluate the effectiveness of the proposed methodology through a series of experiments. For the distillation process, we selected four Teacher models: \emph{Wav2Vec 2.0} base, HuBERT base, WavLM base+, and Robust HuBERT. The distillation process, as outlined in \cite{chang2022distilhubert}, was applied to these models, resulting in models with the \emph{Distiller} prefix. We should note that the distillation process was applied to the {4\textsuperscript{th}, 8\textsuperscript{th}, and 12\textsuperscript{th}} layers of all Teacher models, which may not be optimal for models other than HuBERT \cite{chang2022distilhubert}. For those interested in achieving maximum task-based performance with distilled S3RL, refer to~\cite{app132312571}. Therefore, the reported results may underestimate the achievable accuracy. In the next section, we also show the experimental results of applying our recipe in the DPWavLM method. We follow closely the original procedure described in its original paper.

For all experiments derived from the DistilHuBERT recipe, upstream models were trained using a single NVidia A100 GPU. When utilizing feature denoising KD only, the proposed method requires approximately 30 hours to train, whereas adding the waveform denoising head increases training time to approximately 43 hours. For the experiments based on the DPWavLM recipe, we follow the configurations on the original paper, training with 4 NVidia A100 GPU for approximately 8h. Unless stated otherwise, the reported results were achieved using the feature denoising KD and the STFT-denoising head with the L1 loss. We employed the AdamW optimizer with a batch size of 24 utterances for 200k iterations, and the learning rate linearly decays from $2\times10^{-4}$ to zero after 14k updates.

\subsection{Downstream Tasks, Testing conditions, and Figures-of-Merit}
The ten original tasks from the SUPERB benchmark~\cite{yang2021superb} and two from the SUPERB-SG benchmark~\cite{tsai-etal-2022-superb} are used to gauge the benefits of the proposed method. It is important to emphasize that since our focus is on the upstream model, we do not perform hyperparameter-tuning on the SUPERB downstream tasks; hence we use the default parameters reported in the SUPERB benchmark. Consequently, for the clean test scenario, some of the results reported herein for the baselines may differ from those reported on the SUPERB leaderboard. Nonetheless, the performance trend and gaps reported here are similar to those seen on the leaderboard.

Four evaluation scenarios are considered for analysis for all downstream tasks in our work: {clean} $(c)$, {noise-only} $(n)$, {reverberation-only} $(r)$, and {noise-plus-reverberation} $(n+r)$. The clean scenario meets the same criteria as the test set as SUPERB downstream tasks. For the {noise-only} setting, additive noise with signal-to-noise ratios ranging from $[-5, 20]$ dB are sampled from the DCASE2020 dataset and added to the utterances. For the {reverberation-only} condition, in turn, room impulse responses are uniformly sampled from the OpenSLR26 dataset and convolved with the test set utterance. Lastly, the {noise-plus-reverberation} condition applies both noise and reverberation jointly as described in the $(n)$ and $(r)$ settings, thus representing the most challenging configuration. A fixed custom seed is applied for the random processes to ensure all models are evaluated with the same deterioration.

As for figures-of-merit, each downstream task has its own evaluation metric. For the classification tasks, i.e., KS, SID, IC, and ER, the standard accuracy metric (ACC) is reported. For the SF task, both F-score and Character Error Rate (CER) are used. Word-Error rate (WER), Phoneme-Error rate (PER), and Diarization-Error rate (DER) are the metrics used for the ASR, PR, and SD tasks, respectively. For the ASV task,  the equal-error rate (EER) is used. For the QbE task, we report the maximum term weighted value (MTWV) metric. Next, the SE task is evaluated using quality and intelligibility metrics: the Perceptual Evaluation of Speech Quality (PESQ)~\cite{941023}, the Short-Time Objective Intelligibility (STOI)~\cite{5713237}, and the Scale-Invariant Signal-to-Distortion Ratio (SI-SDR)~\cite{le2019sdr}. Lastly, the SS task is evaluated using the SI-SDR metric as well. More details about the figures-of-merit can be found in \cite{yang2021superb, tsai-etal-2022-superb}.

\section{Experimental Results and Discussion}

\begin{table*}
    \centering
    \caption{Experimental results for content-related tasks on the SUPERB. \textbf{Bold} represents the best result and \underline{underline} the second best for the compressed models.}
    \label{tab:content_results}
    \resizebox{\textwidth}{!}{%
    \begin{tabular}{clcccccccccccccccc}
        \toprule
        & \multirow{2}{*}{Upstream} & \multicolumn{4}{c}{\textbf{KS} - ACC $(\uparrow)$} & \multicolumn{4}{c}{\textbf{PR} - PER $(\downarrow)$} & \multicolumn{4}{c}{\textbf{ASR} - WER $(\downarrow)$} & \multicolumn{4}{c}{\textbf{QbE} - MTWV $(\uparrow)$} \\

        \cmidrule(lr){3-6}
        \cmidrule(lr){7-10}
        \cmidrule(lr){11-14}
        \cmidrule(lr){15-18}

        & & (c) & (n) & (r) & (n + r) & (c) & (n) & (r) & (n + r) & (c) & (n) & (r) & (n + r) & (c) & (n) & (r) & (n + r) \\ 

        \midrule
        & \multicolumn{17}{c}{Teachers} \\
        \midrule
        
        & Wav2Vec 2.0 Base~\cite{baevski2020wav2vec}  & 96.20 & 86.50 & 64.43 & 56.25 & 5.48 & 17.61 & 18.86 & 36.59 & 6.44 & 24.55 & 27.26 & 56.06 & 8.44 & 7.25 & 7.33 & 6.20 \\
        & HuBERT Base~\cite{hsu2021hubert}  & 96.30 & 84.55 & 61.70 & 51.61 & 5.05 & 15.07 & 14.41 & 31.12 & 6.43 & 20.12 & 19.26 & 42.47 & 6.52 & 6.54 & 5.53 & 5.15 \\
        & WavLM Base+~\cite{chen2022wavlm} & 96.88 & 90.07 & 78.64 & 72.38 & 3.93 & 11.33 & 10.83 & 30.63 & 5.87 & 14.36 & 13.94 & 39.17 & 6.25 & 6.02 & 6.16 & 5.29 \\
        & Robust HuBERT~\cite{huang2022improving} & 96.33 & 92.15 & 74.81 & 66.50 & 5.38 & 9.84 & 13.92 & 28.38 & 6.75 & 14.74 & 21.55 & 41.73 & 5.50 & 5.92 & 5.83 & 5.04 \\

        \midrule
        & \multicolumn{17}{c}{Distillation-only recipes} \\
        \midrule
        
        & Distiller (Wav2Vec 2.0)~\cite{chang2022distilhubert} & 95.33 & 82.21 & 55.18 & 45.37 & 15.70 & 37.65 & 67.54 & 77.62 & 13.00 & 46.89 & 113.63 & 181.22 & 6.56 & 4.80 & 5.00 & 5.17 \\
        & Distiller (HuBERT)~\cite{chang2022distilhubert} & 96.14 & 86.34 & 59.20 & 54.53 & 14.15 & 32.95 & 51.40 & 66.87 & 13.26 & 40.55 & 72.07 & 112.73 & \underline{6.72} & 6.14 & 4.99 & 4.44 \\
        & Distiller (WavLM)~\cite{chang2022distilhubert} & 96.36 & 88.48 & 61.34 & 57.58 & \textbf{11.57} & 26.70 & 49.87 & 64.19 & \underline{12.30} & 37.40 & 91.46 & 153.59 & 6.17 & 6.34 & 5.80 & 4.49 \\
        & Distiller (Robust HuBERT)~\cite{chang2022distilhubert} & \underline{96.62} & 87.76 & 58.65 & 50.70 & \underline{12.42} & 28.57 & 56.10 & 71.54 & 12.69 & 36.27 & 91.28 & 145.56 & 6.53 & 6.02 & 4.59 & 4.18 \\
        & FitHuBERT~\cite{lee22p_interspeech} & 96.59 & 89.78 & 61.31 & 56.77 & 13.07 & 27.32 & 38.45 & 55.30 & \textbf{11.56} & 34.20 & 54.54 & 101.33 & 6.21 & 5.79 & 5.06 & 4.38\\


        \hdashline\noalign{\vskip 0.5ex}
        
        & [\textbf{Ours}]~RD~{\footnotesize (Wav2Vec 2.0)} & 96.14 & 91.98 & 85.49 & 78.94 & 16.49 & 24.86 & 26.77 & 38.82 & 13.53 & 25.19 & 27.24 & 43.36 & 6.27 & 6.12 & 5.89 & 4.78 \\
        & [\textbf{Ours}]~RD~{\footnotesize (HuBERT)} & 96.07 & 92.31 & 85.43 & 79.32 & 15.07 & 22.71 & 23.74 & 35.16 & 13.37 & 24.44 & 25.60 & \underline{40.81} & \textbf{6.82} & \textbf{7.35} & \textbf{6.94} & \textbf{6.02}\\ 
        & [\textbf{Ours}]~RD~{\footnotesize (WavLM)} & \textbf{96.79} & \textbf{93.48} & \textbf{87.76} & \textbf{82.31} & 13.38 & \underline{20.68} & \textbf{21.43} & \underline{32.49} & 13.02 & \textbf{23.75} & \textbf{24.46} & \textbf{39.73} & 5.89 & 6.03 & 5.99 & \underline{5.44}\\ 
        & [\textbf{Ours}]~RD~{\footnotesize (Robust HuBERT)} & 96.56 & \underline{92.99} & \underline{85.69} & \underline{79.81} & 13.53 & \textbf{20.58} & \underline{21.50} & \textbf{32.38} & 13.17 & \underline{23.97} & \underline{25.31} & 40.85 &  6.61 & \underline{6.57} & \underline{6.66} & 5.19\\

        \midrule
        & \multicolumn{17}{c}{Distillation-\emph{plus}-Pruning recipes} \\
        \midrule

        & LightHuBERT~\cite{wang2022lighthubert} & 96.11 & 86.69 & \underline{72.35} & \underline{62.25} & \textbf{6.14} & \underline{16.92} & \textbf{12.15} & \underline{27.42} & \textbf{8.22} & \underline{25.33} & \underline{20.30} & \underline{43.80} & 5.90 & \textbf{6.17} & \textbf{6.22} & 4.90 \\
        & DPHuBERT~\cite{peng23c_interspeech} & \underline{96.30} & 84.49 & 63.84 & 56.05 & 9.41 & 24.28 & 29.40 & 49.03 & 10.60 & 32.76 & 40.12 & 76.46 & \textbf{6.28} & 5.91 & 4.96 & \textbf{5.13}\\
        & DPWavLM~\cite{peng23c_interspeech} & 96.20 & \underline{88.12} & 60.40 & 55.08 & \underline{8.23} & 18.81 & 25.95 & 43.78 & 10.14 & 26.17 & 37.20 & 67.93 & \underline{6.04} & 5.46 & 5.20 &\underline{5.05}\\[0.5ex]
        \hdashline\noalign{\vskip 0.5ex}
        
        & [\textbf{Ours}]~RD~{\footnotesize (DPWavLM)} & \textbf{96.69} & \textbf{92.66} & \textbf{88.96} & \textbf{83.93} & 8.26 & \textbf{13.19} & \underline{12.87} & \textbf{21.44} & \underline{9.98} & \textbf{18.37} & \textbf{18.34} & \textbf{31.59} & 6.00 & \underline{6.04} & \underline{5.85} & 5.00\\

        
        \midrule[\heavyrulewidth]
        \bottomrule
    \end{tabular} \vspace{-6mm}
    }
\end{table*}

\subsection{Performance across SUPERB tasks}
In this section, we report the accuracy achieved across the 10 SUPERB tasks with the four baseline Teacher models: Wav2Vec 2.0 base, HuBERT base, WavLM base, and Robust HuBERT, each with 95 million parameters. Moreover, we also report results achieved with different compression recipes, namely \emph{LightHuBERT} (27M parameters), four distilled versions of the Teacher models using the DistilHuBERT recipe (23M parameters, each), the FitHuBERT model (23M parameters), and the DPHuBERT and DPWavLM (24M parameters, each). Finally, we report the performance achieved with the proposed RobustDistiller recipe for the four Teacher models, represented by the acronym \emph{RD} in the tables. The results are shown for the clean test conditions (column `(c) in the tables), noise-only (n), reverberation-only (r), and noise-plus-reverberation (n+r) conditions. The performance metric used by each task is displayed, as well as an up- or down-arrow indicating if higher or lower metric values are better, respectively. Subsection \emph{ 5)} will provide a more detailed discussion of the "distillation-plus-pruning" compression recipe.

\subsubsection{Robustness of Content-related Tasks}
First, we start looking at the robustness experiments, the central part of our contribution. Table~\ref{tab:content_results} shows the obtained results for the four content-related tasks. As can be seen, for the KS task, the \emph{WavLM} model showed the best performance among the Teacher models for three of the four conditions, with Robust HuBERT showing the highest accuracy for the noise-only condition. In the clean and noise-only conditions, the compressed benchmarks performed similarly to the Teacher models; however, the performance degraded substantially under reverberation-only and reverberation-plus-noise settings. The proposed methodology outperformed the original Distiller recipe across all scenarios, even in clean conditions. In the challenging noisy conditions, the RobustDistiller framework allowed for compressed models to even outperform their larger Teacher versions, thus showing the advantage of the feature denoising KD and multi-task learning for speech representation learning and compression.

For the PR and ASR tasks, WavLM and Robust HuBERT continued to show the best accuracy across the Teachers, but with substantial performance drops in the three noisy conditions. For these two tasks, unlike KS, the Distiller compression methodology highly degraded performance with FitHuBERT achieving the top performance on the clean tested conditions. The proposed RobustDistiller framework was able to outperform the original distillation methodology and FitHuBERT on noisy conditions, but not the recipes based on distillation-plus-pruning (e.g., DPWavLM and LightHuBERT). Lastly, for the QbE experiments, the proposed methodology outperformed all the baseline distillation-only schemes and Teacher models, thus further corroborating the importance of the proposed method.

\subsubsection{Robustness of Speaker-related Tasks}

Next, Table~\ref{tab:speaker_results} presents the results for the three speaker-related tasks. For these tasks, reverberation was shown to be a critical nuance factor for the downstream tasks, including the Teacher models, corroborating previous findings \cite{4967982}. For instance, the SID task is based on the VoxCeleb1~\cite{Nagrani17} dataset, a classification problem with more than one thousand classes. Small distribution shifts can substantially affect system performance. Reverberation is known to affect, for example, pitch information, which could be conveyed by the different representations. Of the benchmark compression methods, LightHuBERT and DP-HuBERT-based methods again came out on top for non-robust Students, especially for the conditions involving reverberation. Notwithstanding, the proposed compression recipe achieved the best accuracy in these conditions, often outperforming the respective Teachers.

\begin{table*}
    \centering
    \caption{Experimental results for speaker-related tasks on the SUPERB. \textbf{Bold} represents the best result and \underline{underline} the second best for the compressed models.}
    \label{tab:speaker_results}
    \resizebox{\textwidth}{!}{%
    \begin{tabular}{clccccccccccccc}
        \toprule
        & \multirow{2}{*}{Upstream} & \multicolumn{4}{c}{\textbf{SID} - ACC $(\uparrow)$} & \multicolumn{4}{c}{\textbf{ASV} - EER $(\downarrow)$} & \multicolumn{4}{c}{\textbf{SD} - DER $(\downarrow)$} \\

        \cmidrule(lr){3-6}
        \cmidrule(lr){7-10}
        \cmidrule(lr){11-14}

        & & (c) & (n) & (r) & (n + r) & (c) & (n) & (r) & (n + r) & (c) & (n) & (r) & (n + r) \\ 

        \midrule
        & \multicolumn{13}{c}{Teachers} \\
        \midrule
        & Wav2Vec 2.0 Base~\cite{baevski2020wav2vec} & 75.17 & 41.79 & 5.36 & 2.80 & 6.01 & 15.33 & 28.45 & 34.05 & 6.08 & 8.33 & 40.44 & 46.91 \\
        & HuBERT Base~\cite{hsu2021hubert} & 81.43 & 52.44 & 12.37 & 6.46 & 5.50 & 12.64 & 24.41 & 28.80 & 5.88 & 8.02 & 28.51 & 32.55 \\
        & WavLM Base+~\cite{chen2022wavlm} & 72.14 & 49.52 & 14.08 & 6.74 & 4.34 & 10.16 & 20.88 & 28.90 & 3.58 & 5.50 & 16.25 & 21.11 \\
        & Robust HuBERT~\cite{huang2022improving} & 73.08 & 48.43 & 6.99 & 4.24 & 5.26 & 10.05 & 26.01 & 31.28 & 4.79 & 6.21 & 19.51 & 22.49 \\
        
        \midrule
        & \multicolumn{13}{c}{Distillation-only recipes} \\
        \midrule

        & Distiller (Wav2Vec 2.0)~\cite{chang2022distilhubert} & 73.63 & 41.17 & 3.07 & 1.49 & 7.27 & 17.35 & 39.18 & 41.90 & 6.59 & 8.31 & 59.72 & 65.24 \\
        & Distiller (HuBERT)~\cite{chang2022distilhubert} & 72.28 & 45.01 & 5.11 & 2.95 & 6.59 & 13.94 & 29.32 & 33.84 & 5.98 & 7.95 & 39.51 & 41.98 \\
        & Distiller (WavLM)~\cite{chang2022distilhubert} & \textbf{79.78} & 52.19 & 6.07 & 3.51 & \textbf{5.69} & 11.51 & 33.02 & 35.82 & 5.75 & 7.74 & 53.52 & 60.11 \\
        & Distiller (Robust HuBERT)~\cite{chang2022distilhubert} & 77.37 & 50.67 & 5.15 & 3.07 & 6.08 & 12.08 & 36.90 & 40.01 & 6.12 & 7.71 & 42.67 & 49.27 \\
        & FitHuBERT~\cite{lee22p_interspeech} & 54.09 & 33.04 & 7.22 & 4.04 & 6.72 & 13.86 & 33.49 & 37.75 & 6.84 & 8.99 & 33.79 & 40.63 \\
        
        \hdashline\noalign{\vskip 0.5ex}
        
        &  [\textbf{Ours}]~RD~{\footnotesize (Wav2Vec 2.0)} & 76.27 & 61.06 & 40.18 & 28.86 & 7.31 & 10.69 & 13.51 & 16.89 & 5.22 & 6.99 & 7.69 & \textbf{9.39} \\
        & [\textbf{Ours}]~RD~{\footnotesize (HuBERT)} & 77.91 & \underline{61.83} & \textbf{43.56} & \textbf{31.66} & 6.23 & \underline{9.23} & \underline{11.66} & \underline{15.01} & \underline{5.19} & \underline{6.40} & \textbf{8.01} & \underline{10.46} \\
        & [\textbf{Ours}]~RD~{\footnotesize (WavLM)} &  \underline{78.39} & 60.73 & \underline{42.04} & \underline{30.02} & \underline{5.83} & \textbf{8.93} & \textbf{11.24} & \textbf{14.59} & \textbf{4.92} & \textbf{6.28} & 9.01 & 12.09 \\
        & [\textbf{Ours}]~RD~{\footnotesize (Robust HuBERT)} & 77.52 & \textbf{61.96} & 41.79 & 29.66 & 6.59 & 9.64 & 12.22 & 16.02 & 5.31 & 6.53 & \underline{8.15} & 10.63 \\
        
        \midrule
        & \multicolumn{13}{c}{Distillation-\emph{plus}-Pruning recipes} \\
        \midrule
        & LightHuBERT~\cite{wang2022lighthubert} & 72.03 & 41.57 & \underline{17.50} & \underline{8.68} & 5.57 & 12.80 & \underline{14.81} & \underline{23.56} & 6.08 & 7.96 & \underline{19.38} & \underline{21.73} \\
        & DPHuBERT~\cite{peng23c_interspeech} & 76.33 & 45.79 & 8.13 & 4.27 & \underline{5.54} & 12.93 & 23.93 & 30.80 & 6.16 & 7.97 & 26.18 & 32.08 \\
        & DPWavLM~\cite{peng23c_interspeech} & \underline{82.18} & \underline{53.65} & 12.48 & 6.61 & 5.58 & \underline{10.66} & 29.69 & 34.84 & \underline{5.56} & \underline{7.18} & 36.53 & 42.88 \\
        
        \hdashline\noalign{\vskip 0.5ex}
        
        & [\textbf{Ours}]~RD~{\footnotesize (DPWavLM)} & \textbf{84.01} & \textbf{67.69} & \textbf{49.63} & \textbf{35.06} & \textbf{5.05} & \textbf{8.14} & \textbf{9.89} & \textbf{13.99} & \textbf{5.08} & \textbf{6.37} & \textbf{6.64} & \textbf{8.45} \\
        
        \midrule[\heavyrulewidth]
        \bottomrule
    \end{tabular} 
    }
\end{table*}

\subsubsection{Robustness of Semantic-related Tasks}

For the semantic-related evaluations, Table \ref{tab:semantics_results} shows the results for two tasks: IC and SF. Baseline compression highly affected system performance in noisy conditions, with distillation-plus-pruning recipes again showing improved accuracy across all tested conditions. The proposed RobustDistiller consistently outperformed the original Distiller and FitHuBERT recipes in several cases, achieving results in line with LightHuBERT, DPHuBERT, or even surpassing the Teacher model (e.g., in the IC task under the n+r condition). 

\begin{table*}
    \centering
    \caption{Experimental results for semantics-related tasks on the SUPERB. \textbf{Bold} represents the best result and \underline{underline} the second best for the compressed models.}
    \label{tab:semantics_results}
    \resizebox{\textwidth}{!}{%
    \begin{tabular}{clccccccccccccc}
        \toprule
        & \multirow{2}{*}{Upstream} & \multicolumn{4}{c}{\textbf{IC} - ACC $(\uparrow)$} & \multicolumn{4}{c}{\textbf{SF} - F1 $(\uparrow)$} & \multicolumn{4}{c}{\textbf{SF} - CER $(\downarrow)$} \\

        \cmidrule(lr){3-6}
        \cmidrule(lr){7-10}
        \cmidrule(lr){11-14}

        & & (c) & (n) & (r) & (n + r) & (c) & (n) & (r) & (n + r) & (c) & (n) & (r) & (n + r) \\ 

        \midrule
        & \multicolumn{13}{c}{Teachers} \\
        \midrule
 
        & Wav2Vec 2.0 Base~\cite{baevski2020wav2vec} & 92.83 & 69.26 & 68.78 & 50.88 & 88.30 & 74.06 & 70.20 & 47.61 & 24.77 & 41.73 & 48.47 & 68.11 \\
        & HuBERT Base~\cite{hsu2021hubert} & 98.34 & 78.70 & 75.77 & 59.08 & 88.52 & 76.90 & 77.16 & 57.37 & 25.20 & 38.98 & 40.92 & 59.58 \\
        & WavLM Base+~\cite{chen2022wavlm} & 98.84 & 83.13 & 89.35 & 68.20 & 90.14 & 83.74 & 85.62 & 67.73 & 21.92 & 30.08 & 29.71 & 49.06 \\
        & Robust HuBERT~\cite{huang2022improving} & 98.66 & 91.43 & 85.32 & 71.10 & 88.83 & 84.14 & 78.75 & 61.75 & 24.24 & 31.25 & 38.61 & 55.69 \\

        \midrule
        & \multicolumn{13}{c}{Distillation-only recipes} \\
        \midrule
        
        & Distiller (Wav2Vec 2.0)~\cite{chang2022distilhubert} & 86.05 & 39.81 & 29.24 & 17.37 & 82.28 & 60.58 & 25.22 & 16.13 & 35.88 & 59.88 & 88.42 & 93.66 \\
        & Distiller (HuBERT)~\cite{chang2022distilhubert} & 94.12 & 48.72 & 47.72 & 24.31 & 82.12 & 64.14 & 34.43 & 20.83 & 35.21 & 55.44 & 82.26 & 90.56 \\
        & Distiller (WavLM)~\cite{chang2022distilhubert} & 94.89 & 63.25 & 48.43 & 31.69 & \underline{84.76} & 68.74 & 36.72 & 21.43 & 32.89 & 50.60 & 80.47 & 89.66 \\
        & Distiller (Robust HuBERT)~\cite{chang2022distilhubert} & \textbf{95.86} & 67.07 & 46.30 & 28.34 & \textbf{85.19} & 68.71 & 29.08 & 17.06 & \textbf{30.98} & 49.69 & 85.84 & 92.92 \\
        & FitHuBERT~\cite{lee22p_interspeech} & 94.89 & 63.56 & 67.20 & 45.48 & 84.75 & 70.68 & 59.90 & 39.04 & \underline{31.03} & 46.97 & 59.73 & 76.47 \\
        
        \hdashline\noalign{\vskip 0.5ex}

        & [\textbf{Ours}]~RD~{\footnotesize (Wav2Vec 2.0)} & 89.30 & 71.47 & 71.00 & 54.47 & 80.61 & 73.68 & 72.12 & 60.32 & 37.58 & 46.92 & 49.71 & 62.51 \\
        & [\textbf{Ours}]~RD~{\footnotesize (HuBERT)} & 93.78 & 78.96 & 78.01 & \underline{66.28} & 82.26 & 75.04 & 74.45 & 61.97 & 35.70 & 45.24 & \underline{46.65} & 59.96 \\
        & [\textbf{Ours}]~RD~{\footnotesize (WavLM)} &  95.23 & \textbf{82.05} & \textbf{86.45} & \textbf{72.21} & 82.46 & \textbf{75.93} & \textbf{75.38} & \textbf{62.88} & 35.88 & \underline{43.94} & \textbf{45.85} & \textbf{58.55} \\
        & [\textbf{Ours}]~RD~{\footnotesize (Robust HuBERT)} & \underline{95.81} & \underline{81.12} & \underline{84.16} & 64.59 & 83.27 & \underline{75.81} & \underline{74.89} & \underline{62.44} & 34.42 & \textbf{43.73} & 46.72 & \underline{58.82} \\

        \midrule
        & \multicolumn{13}{c}{Distillation-\emph{plus}-Pruning recipes} \\
        \midrule

        & LightHuBERT~\cite{wang2022lighthubert} & \textbf{98.60} & 83.73 & \underline{88.19} & \underline{70.45} & 87.70 & 74.82 & \underline{79.78} & \underline{60.67} & \underline{26.04} & 42.24 & \underline{38.57} & \underline{58.33} \\
        & DPHuBERT~\cite{peng23c_interspeech} & 97.89 & 73.45 & 72.29 & 50.25 & 87.41 & 72.91 & 68.74 & 44.37 & 28.36 & 44.85 & 52.17 & 72.89 \\
        & DPWavLM~\cite{peng23c_interspeech} & \underline{98.58} & \underline{83.94} & 77.72 & 61.24 & \underline{88.03} & \underline{78.25} & 75.05 & 55.23 & \textbf{25.93} & \underline{38.72} & 44.78 & 63.15 \\[0.5ex]
        
        \hdashline\noalign{\vskip 0.5ex}

        & [\textbf{Ours}]~RD~{\footnotesize (DPWavLM)} & 98.39 & \textbf{92.33} & \textbf{92.67} & \textbf{83.18} & \textbf{88.21} & \textbf{82.69} & \textbf{84.03} & \textbf{72.69} & 26.80 & \textbf{34.84} & \textbf{34.67} & \textbf{47.72}  \\

        \midrule[\heavyrulewidth]
        \bottomrule
    \end{tabular} 
    }
\end{table*}

\subsubsection{Robustness of Paralinguistic-related Tasks}

Finally, for the ER task, Table~\ref{tab:paralinguistic_results} summarizes our findings. Compression was shown to be fairly resilient to the noisy conditions and resulted in recognition accuracy in line with those obtained with the Teacher models. The RobustDistiller method achieved the highest accuracy of all tested distillation-only compression models and was the closest to the accuracy of the Teacher models, despite requiring four times fewer parameters. FitHuBERT showed the best accuracy on the noise-only scenario among the compressed models.

\begin{table}
    \centering
    \caption{Experimental results for the paralinguistic task (emotion recognition) on the SUPERB. \textbf{Bold} represents the best result and \underline{underline} the second best for the compressed models.}
    \label{tab:paralinguistic_results}
    \resizebox{0.5\textwidth}{!}{%
    \begin{tabular}{clcccc}
        \toprule
        & \multirow{2}{*}{Upstream} & \multicolumn{4}{c}{\textbf{ER} - ACC $\uparrow$} \\

        \cmidrule(lr){3-6}

        & & (c) & (n) & (r) & (n + r) \\

        \midrule
        & \multicolumn{5}{c}{Teachers} \\
        \midrule
 
        & Wav2Vec 2.0 Base~\cite{baevski2020wav2vec} & 63.01 & 51.46 & 32.70 & 27.02 \\
        & HuBERT Base~\cite{hsu2021hubert} &  64.75 & 53.45 & 40.72 & 29.12 \\
        & WavLM Base+~\cite{chen2022wavlm} & 67.96 & 55.08 & 36.57 & 27.14 \\
        & Robust HuBERT~\cite{huang2022improving} & 64.59 & 57.17 & 41.44 & 33.36 \\
        
        \midrule
        & \multicolumn{5}{c}{Distillation-only recipes} \\
        \midrule
        
        & Distiller (Wav2Vec 2.0)~\cite{chang2022distilhubert} &  61.01 & 49.94 & 24.14 & 23.69 \\
        & Distiller (HuBERT)~\cite{chang2022distilhubert} & 62.55 & 47.89 & 31.35 & 25.77 \\
        & Distiller (WavLM)~\cite{chang2022distilhubert} & 62.09 & 51.14 & 29.24 & 25.27 \\
        & Distiller (Robust HuBERT)~\cite{chang2022distilhubert} & \textbf{64.02} & 51.60 & 29.29 & 25.08 \\
        & FitHuBERT~\cite{lee22p_interspeech} & 61.35 & \textbf{52.17} & 36.25 & 28.06 \\
        
        \hdashline\noalign{\vskip 0.5ex}
        
        & [\textbf{Ours}]~RD~{\footnotesize (Wav2Vec 2.0)} & \underline{63.57} & 50.50 & \textbf{40.08} & \underline{29.98} \\
        & [\textbf{Ours}]~RD~{\footnotesize (HuBERT)} & 61.46 & 48.06 & 35.40 & 27.28 \\
        & [\textbf{Ours}]~RD~{\footnotesize (WavLM)} &  62.76 & 51.50 & \underline{39.73} & \textbf{32.65} \\
        & [\textbf{Ours}]~RD~{\footnotesize (Robust HuBERT)} &  62.81 & \underline{52.10} & 34.72 & 27.36 \\

        \midrule
        & \multicolumn{5}{c}{Distillation-\emph{plus}-Pruning recipes} \\
        \midrule
        
        & LightHuBERT~\cite{wang2022lighthubert} & 63.88 & 48.72 & 35.51 & 29.84 \\
        & DPHuBERT~\cite{peng23c_interspeech} & 63.06 & 52.90 & 35.45 & 27.44 \\
        & DPWavLM~\cite{peng23c_interspeech} & \textbf{65.61} & \underline{57.58} & \underline{36.87} & \underline{29.96} \\[0.5ex]
        
        \hdashline\noalign{\vskip 0.5ex}

        & [\textbf{Ours}]~RD~{\footnotesize (DPWavLM)} & \underline{64.87} & \textbf{61.10} & \textbf{59.78} & \textbf{49.04} \\

        \midrule[\heavyrulewidth]
        \bottomrule
    \end{tabular} 
    }
\end{table}

\subsubsection{Generalizing to other Teacher-Student recipes}

The last subgroup in Tables~\ref{tab:content_results} to \ref{tab:paralinguistic_results} describes the results obtained from models that use distillation-plus-pruning recipes. Empirically, these models achieve the best overall performance among the compressed models, especially on content-related tasks (e.g., ASR). Beyond a similar generalization to its Teacher on clean scenarios, these models also exhibit some degree of robustness to adverse conditions. For instance, the LightHuBERT model is one of the most resilient to the reverberation condition, sometimes even outperforming its Teacher, the HuBERT model. In this work, to show how flexible our recipe is, we selected the DPWavLM model due to its excellent generalization capabilities and feasibility for training under academic resources. DPWavLM consists of two stages. First, distillation and pruning are jointly performed. In the second stage, only distillation is used to adjust the Student. We specifically insert our two modifications of the RobustDistiller in the second stage.

Our results show that the RobustDistiller recipe can increase the robustness of the Student model and help with the generalization for clean scenarios, as observed in several tasks such as ASR, KS, SID, and SD. For the noisy scenarios, the gap between clean and noisy settings is significantly smaller. For instance, the RobustDistiller on top of the DPWavLM in the noise-plus-reverberation scenario for the IC task has a drop of 15.21\% percentage points in accuracy. In contrast, Robust HuBERT, a robust model four times the size of our Student, has a drop of 27.56\%.

\subsubsection{Overall Performance across tasks and scenarios}

Lastly, Table~\ref{tab:final_scores_results} shows the overall score (Eq. 1) grouped by task type and evaluation scenario. We also present the number of parameters and the number of multiply-accumulate operations (MACs) for each model. As expected for models tested under clean settings, the Teacher models have better overall performance when compared to their smaller counterparts. For the compressed models, we observe that distillation-plus-pruning takes the lead across clean tasks, with a commendable performance for DPWavLM and LightHuBERT. Nonetheless, the proposed RobustDistiller shows that it can not only improve ``in-the-wild'' scenarios, but also help the model generalize better for clean settings. For instance, for content- and speaker-related tasks, DPWavLM trained with the proposed recipe improves the overall performance under clean scenarios, with 17.15\% and 20.22\% increase in the score, respectively. Speaker-related tasks, in general, greatly benefit from the proposed robust distillation recipe for both distillation-plus-pruning and distillation-only methods.

For noisy conditions, we observe that RobustDistiller can drastically improve across all four groups of tasks, sometimes even surpassing robust Teacher models, such as WavLM and Robust HuBERT. Comparing non-robust distillation-only methods with the proposed RobustDistiller on semantics-related tasks with reverberation, we can observe a Score gain of 2.6 times. Similar results can be observed on speaker-based tasks. Overall, the gains achieved with RobustDistiller motivate the development of distillation methods with a robustness-first principle during its development.

\begin{table*}
    \centering
    \caption{Overall Score results for each scenario on the original 10 SUPERB tasks grouped by category. \textbf{Bold} represents the best result and \underline{underline} the second best for the compressed models.}
    \label{tab:final_scores_results}
    \resizebox{\textwidth}{!}{%
    {\begin{tabular}{clccccccccccccccccccccccccc}
        \toprule
        & \multirow{2}{*}{Upstream} & \#params & MACs & \multicolumn{4}{c}{Content-related Tasks} & \multicolumn{4}{c}{Speaker-related Tasks}  & \multicolumn{4}{c}{Semantic-related Tasks} & \multicolumn{4}{c}{Paralinguistics-related Tasks} \\
        \cmidrule(lr){5-8}
        \cmidrule(lr){9-12}
        \cmidrule(lr){13-16}
        \cmidrule(lr){17-20}

        & & (M) & (G) & (c) & (n) & (r) & (n + r) & (c) & (n) & (r) & (n + r) & (c) & (n) & (r) & (n + r) & (c) & (n) & (r) & (n + r) \\

        \midrule
        & \multicolumn{20}{c}{Teachers} \\
        \midrule

        & Wav2Vec 2.0 Base~\cite{baevski2020wav2vec} & 95 & 1669 & 841 & 686 & 750 & 712 & 459 & 221 & 260 & 214 & 671 & 576 & 668 & 533 & 289 & 270 & 240 & 131 \\
        & HuBERT Base~\cite{hsu2021hubert} & 95 & 1669 &  702 & 630 & 605 & 600 & 605 & 450 & 431 & 398 & 885 & 719 & 784 & 685 & 539 & 421 & 465 & 214 \\
        & WavLM Base+~\cite{chen2022wavlm} & 95 & 1670 & 814 & 780 & 817 & 759 & 868 & 752 & 560 & 466 & 1000 & 908 & 974 & 857 & 1000 & 545 & 349 & 136 \\
        & Robust HuBERT~\cite{huang2022improving} & 95 & 1669 & 604 & 827 & 726 & 697 & 651 & 678 & 431 & 405 & 921 & 982 & 876 & 817 & 516 & 703 & 485 & 382 \\
        
        \midrule
        & \multicolumn{20}{c}{Distillation-only recipes} \\
        \midrule
        
        & Distiller (Wav2Vec 2.0)~\cite{chang2022distilhubert} & 23 & 786 & 123 & 0 & 37 & 123 & 248 & 143 & 0 & 0 & 71 & 0 & 0 & 0 & 0 & 155 & 0 & 0 \\
        & Distiller (HuBERT)~\cite{chang2022distilhubert} & 23 & 786 & 289 & 314 & 242 & 254 & 371 & 338 & 254 & 247 & 393 & 160 & 210 & 90 & 222 & 0 & 202 & 82 \\
        & Distiller (WavLM)~\cite{chang2022distilhubert} & 23 & 786 & 362 & 461 & 290 & 224 & \underline{579} & 515 & 131 & 123 & 529 & 388 & 233 & 154 & 156 & 246 & 143 & 62 \\
        & Distiller (Robust HuBERT)~\cite{chang2022distilhubert} & 23 & 786 & \textbf{405} & 405 & 132 & 121 & 471 & 483 & 148 & 132 & \textbf{609} & 431 & 161 & 91 & \textbf{433} & 281 & 145 & 55 \\
        & FitHuBERT~\cite{lee22p_interspeech} & 22 & 3122 & \underline{395} & 455 & 365 & 331 & 67 & 126 & 257 & 219 & \underline{559} & 442 & 565 & 408 & 50 & \underline{324} & 340 & 172 \\[0.5ex]

        \hdashline\noalign{\vskip 0.5ex}       
        
        & [\textbf{Ours}]~RD~{\footnotesize (Wav2Vec 2.0)} & 23 & 786 & 195 & 628 & 740 & 695 & 413 & 702 & 885 & 898 & 127 & 549 & 688 & 647 & \underline{368} & 198 & \textbf{447} & \underline{248} \\
        & [\textbf{Ours}]~RD~{\footnotesize (HuBERT)} & 23 & 786 & 265 & \textbf{781} & \textbf{851} & \textbf{871} & 555 & \underline{819} & \textbf{928} & \textbf{942} & 375 & 649 & 766 & \underline{758} & 65 & 13 & 316 & 141 \\
        & [\textbf{Ours}]~RD~{\footnotesize (WavLM)} & 23 & 786 & 346 & 700 & 795 & \underline{833} & \textbf{633} & \textbf{830} & \underline{915} & \underline{921} & 434 & \textbf{699} & \textbf{840} & \textbf{814} & 253 & 273 & \underline{437} & \textbf{354} \\
        & [\textbf{Ours}]~RD~{\footnotesize (Robust HuBERT)} & 23 & 786 & 363 & \underline{742} & \underline{839} & 785 & 499 & 792 & 908 & 909 & 502 & \underline{690} & \underline{816} & 753 & 259 & \textbf{319} & 297 & 145 \\

        \midrule
        & \multicolumn{20}{c}{Distillation-\emph{plus}-Pruning recipes} \\
        \midrule

        & LightHuBERT~\cite{wang2022lighthubert} & 27 & 861 & \underline{538} & \underline{585} & \underline{754} & \underline{652} & 473 & 345 & \underline{634} & \underline{546} & \underline{861} & 717 & \underline{903} & \underline{792} & 413 & 63 & 319 & 243 \\
        & DPHuBERT~\cite{peng23c_interspeech} & 24 & 654 & 459 & 387 & 451 & 489 & 516 & 380 & 420 & 355 & 788 & 577 & 674 & 488 & 295 & 379 & 317 & 148 \\
        & DPWavLM~\cite{peng23c_interspeech} & 24 & 589 & 462 & 524 & 470 & 512 & \underline{638} & \underline{614} & 321 & 266 & \textbf{870} & \underline{785} & 774 & 672 & \textbf{662} & \underline{734} & \underline{357} & \underline{247} \\[0.5ex]
        
        \hdashline\noalign{\vskip 0.5ex}

        & [\textbf{Ours}]~RD~{\footnotesize (DPWavLM)} & 24 & 589 & \textbf{541} & \textbf{792} & \textbf{845} & \textbf{852} & \textbf{766} & \textbf{917} & \textbf{1000} & \textbf{1000} & 854 & \textbf{945} & \textbf{972} & \textbf{1000} & \underline{556} & \textbf{1000} & \textbf{1000} & \textbf{1000} \\
        
        \midrule[\heavyrulewidth]
        \bottomrule
    \end{tabular}} 
    }
\end{table*}

\subsection{Performance on SUPERB-SG Tasks: Speech Enhancement and Source Separation}

For downstream generative tasks, our assumption is that feature denoising KD and multi-task learning will lead to improved disentanglement in the latent space, resulting in improved results. Table~\ref{tab:superbsg_results} reports the results obtained for the SE and SS tasks. One of the first observations for these tasks is that compressed models do not have a significant performance gap compared to their Teacher model, as has happened on other tasks. For distillation-only recipes, the  proposed modifications led to improvements in both SE and SS compared to their original recipe. For distillation-plus-pruning recipes, however, there is no clear pattern on whether RobustDistiller helps or not. Furthermore, examining the overall results for smaller models, robust or not, these experiments suggest that there is still room for improvements on generative tasks.

\begin{table}
    \centering
    \caption{Experimental results on the SUPERB-SG benchmark for Speech Enhancement (SE) and Source Separation (SS) tasks. \textbf{Bold} represents the best result and \underline{underline} the second best for the compressed models.}
    \label{tab:superbsg_results}
    \resizebox{0.5\textwidth}{!}{%
    \begin{tabular}{clccccc}
        \toprule
        & \multirow{2}{*}{Upstream} & \multicolumn{3}{c}{SE} & SS & \multirow{2}{*}{Score $(\uparrow)$} \\
        \cmidrule(lr){3-5}
        \cmidrule(lr){6-6}
        & & STOI $(\uparrow)$ & PESQ $(\uparrow)$ & SI-SDR $(\uparrow)$ & SI-SDR $(\uparrow)$ & \\
        \midrule
        & \multicolumn{6}{c}{Teachers} \\
        \midrule

        & Wav2Vec 2.0 Base~\cite{baevski2020wav2vec} & 94.83 & 2.93 & 9.34 & 10.02 & 715.07 \\
        & HuBERT Base~\cite{hsu2021hubert} & 94.86 & 2.98 & 9.05 & 9.97 & 726.45  \\
        & WavLM Base+~\cite{chen2022wavlm} & 95.17 & 3.04 & 9.41 & 11.37 & 988.09 \\
        & Robust HuBERT~\cite{huang2022improving}  & 94.93 & 3.00 & 9.16 & 9.86 & 751.27  \\

        \midrule
        & \multicolumn{6}{c}{Distillation-only recipes} \\
        \midrule

        & Distiller (Wav2Vec 2.0)~\cite{chang2022distilhubert} & 94.71 & 2.91 & 9.41 & 9.48 & 633.51 \\
        & Distiller (HuBERT)~\cite{chang2022distilhubert} & 94.76 & 2.96 & 9.24 & 9.39 & 656.65  \\
        & Distiller (WavLM)~\cite{chang2022distilhubert} & 94.88 & 2.98 & 8.71 & \underline{10.03} & 702.04 \\
        & Distiller (Robust HuBERT)~\cite{chang2022distilhubert} & 94.85 & 2.94 & 9.44 & 9.32 & 671.14 \\
        & FitHuBERT~\cite{lee22p_interspeech} & 94.31 & 2.79 & 7.75 & 6.02 & 0.00 \\

        \hdashline\noalign{\vskip 0.5ex}
        & [\textbf{Ours}]~RD~{\footnotesize (Wav2Vec 2.0)} & 94.89 & \underline{2.99} & 9.19 & 10.01 & 752.93  \\
        & [\textbf{Ours}]~RD~{\footnotesize (HuBERT)} & \underline{94.90} & \textbf{3.00} & \textbf{9.47} & 9.80 & \underline{770.31}  \\
        & [\textbf{Ours}]~RD~{\footnotesize (WavLM)} & \textbf{94.95} & \underline{2.99} & \underline{9.46} & \textbf{10.17} & \textbf{802.38} \\
        & [\textbf{Ours}]~RD~{\footnotesize (Robust HuBERT)} & 94.82 & \underline{2.99} & 9.14 & 9.73 & 711.64 \\

        \midrule
        & \multicolumn{6}{c}{Distillation-\emph{plus}-Pruning recipes} \\
        \midrule
        & LightHuBERT~\cite{wang2022lighthubert} & 94.72 & 2.96 & 9.11 & 9.91 & 683.06 \\
        & DPHuBERT~\cite{peng23c_interspeech} & 94.81 & 2.92 & 9.09 & 9.80 & 662.85 \\
        & DPWavLM~\cite{peng23c_interspeech} & \underline{94.92} & \textbf{3.02} & \underline{9.52} & \textbf{10.56} & \textbf{861.84} \\[0.5ex]
        
        \hdashline\noalign{\vskip 0.5ex}
        & [\textbf{Ours}]~RD~{\footnotesize (DPWavLM)} & \textbf{94.95} & \underline{3.00} & \textbf{9.53} & \underline{10.47} & \underline{846.80} \\

        \midrule[\heavyrulewidth]
        \bottomrule
    \end{tabular}
    }
\end{table}

\subsection{Speech-noise disentanglement}
\begin{figure}
        \centering
        \includegraphics[width=\linewidth]{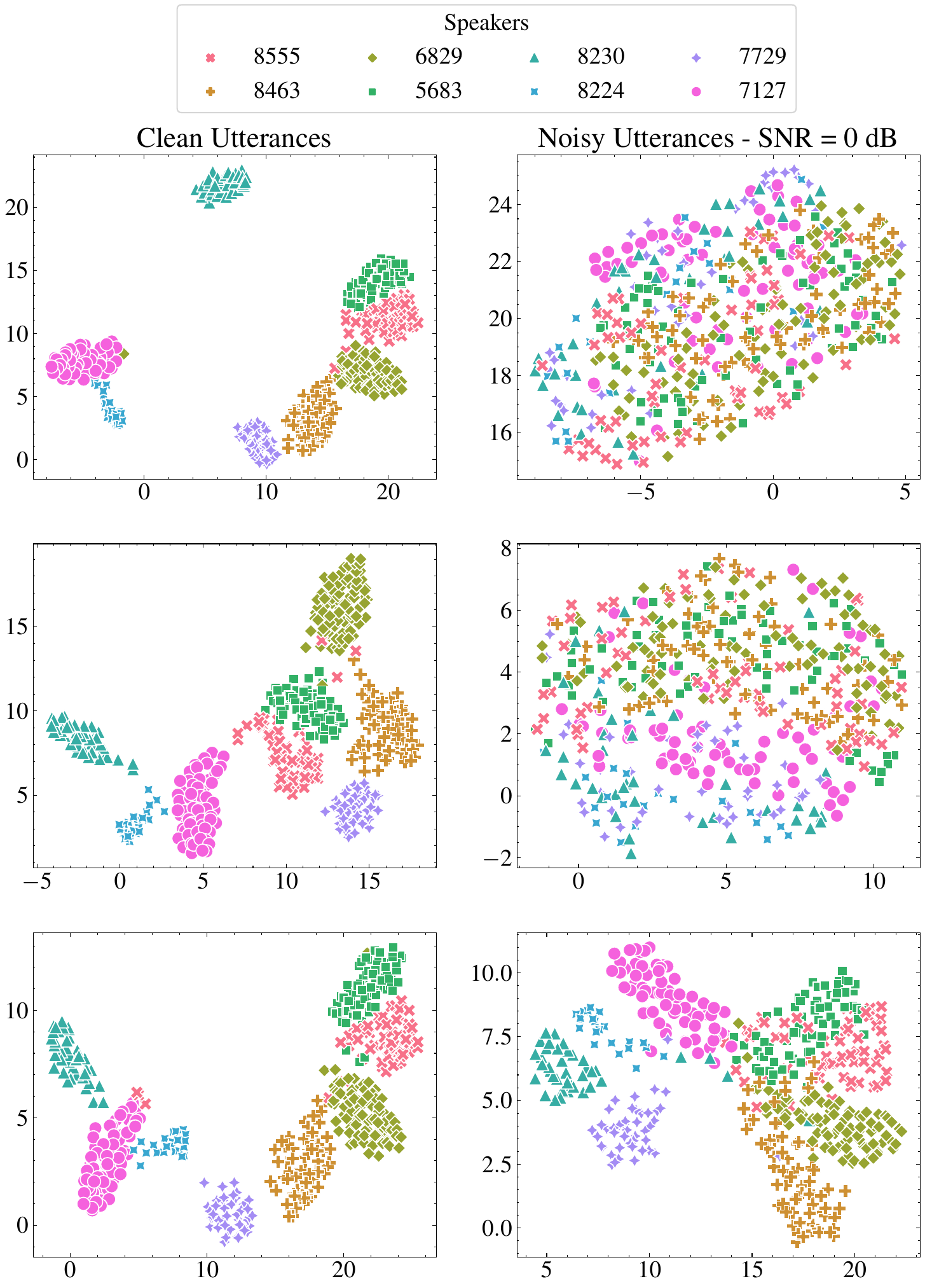} \caption{tSNE visualization on the effect of noise on the latent representation of (top to bottom) HuBERT, Distiller (HuBERT), and RobustDistiller (HuBERT), respectively. Clean signals are shown on the left, while noisy on the right. Data are from eight randomly-selected speakers of LibriSpeech.}
        \label{fig:spk_latent_rep_noise}
\end{figure}

The effects of feature denoising knowledge distillation and the multi-task learning during distillation proved especially useful in clean and noisy settings. Initially, we hypothesized that both paradigms helped the model to better generalize by reinforcing the disentanglement of speech features from noise factors. In Figure~\ref{fig:spk_latent_rep_noise}, we further explore this hypothesis and test the effects that noise has on the latent space of three different models: HuBERT, Distiller (HuBERT), and RobustDistiller (HuBERT), from top to bottom, respectively. Left-side plots are for the clean utterances and right-side for noise under a 0 dB SNR setting. Each individual point is an utterance, and its shape/color represents one out of eight randomly-selected speakers from the LibriSpeech dataset. First, we compute a global average pooling of each frame of the tested representation. Then we reduce the dimensionality of the signal to two dimensions via a tSNE transformation. As can be seen, the representations are able to accurately discriminate the speakers in the clean condition. In turn, under the noisy setting, only the RobustDistiller method preserves the disentanglement across speakers.

\subsection{Impact of the Enhancement Head}

\begin{figure*}
   \centering
   \includegraphics[width=\linewidth]{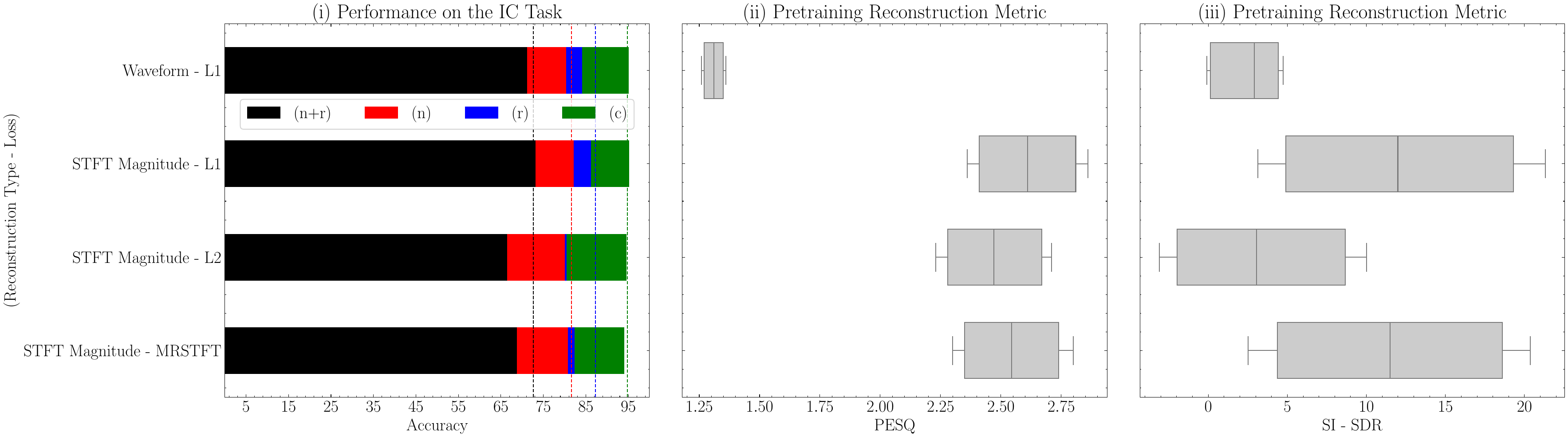}
   \caption{Measuring the impact of different enhancement heads and loss functions for our proposed multi-task learning paradigm.}
   \label{fig:enhc_exp}
\end{figure*}

To evaluate the effectiveness of the proposed modification \#2 (i.e., adding the enhancement head), we conducted experiments with models capable of reconstructing either the original waveform or the clean STFT magnitude, using different enhancement losses. During the distillation phase, we measured reconstruction metrics (e.g., PESQ and SI-SDR) every 50 steps for each model configuration. Once the distilled model was obtained, we fine-tuned it for three different tasks: KS, IC, and ER.

Figure~\ref{fig:enhc_exp} (i) depicts the results for the IC task under the four tested scenarios. The dashed lines indicate the outcome for a model without the enhancement head but which has the feature denoising knowledge distillation step. Figure~\ref{fig:enhc_exp} (ii) and (iii) display the pretraining PESQ and SI-SDR, respectively.

As can be seen, the combination that achieves the best overall performance relies on an enhancement head that estimates the binary mask for the STFT magnitude and uses an L1-loss function. This result highly correlates with the best reconstruction metrics observed during the pretraining phase. We hypothesize that, for small-capacity enhancement heads that estimate a binary mask, the greater the reconstruction during the training, the better the final disentanglement because the Student model is learning to preserve more speech features in their latent space. In particular, for the IC task in Figure~\ref{fig:enhc_exp}(i), we observe an improvement in three of the four tested scenarios using the enhancement head. For the reverberation scenario, the usage of our proposed enhancement head slightly improved upon the setting where we use only feature denoising KD; thus, a limitation of this work is to experiment with better architectures that can handle better dereverberation.

Despite achieving poor reconstruction metrics during pretraining, the second-best overall result across the three tasks goes for the model reconstructing the raw waveform directly and using an L1-loss function. Going from a very compressed latent representation to a high-dimensional signal (i.e., raw waveform) is difficult. Furthermore, inserting a more robust enhancement head can soften the job from the Student model, thus undermining the full potential of creating an encoder that disentangles noise from speech features as we planned.

\subsection{Discussion on distilling from already robust Teachers}
One important question that may arise from the analysis herein is the following: What if we distill a representation that was built to be robust; do we get better results than developing a robust distillation process? To answer this question, we zoom into the results in Table~\ref{tab:final_scores_results}. In this analysis, we consider the WavLM and Robust HuBERT models to be robust Teachers since both were augmented with noise during the training phase. First, considering distillation-only recipes, we observe that the Distiller recipe with HuBERT and Robust HuBERT leads to the former Student being more robust in the noise-only setting. However, both have a huge gap compared to the Teacher model. When applying our methodology, RobustDistiller, we observe a large improvement in all three noisy scenarios.

Interestingly, RobustDistiller with HuBERT or the Robust HuBERT models generate Students with similar performance. Thus, a two-step process of first making the Teacher model and then distilling it is not necessary. It is computationally better to distill a non-robust model directly with RobustDistiller, and we can achieve similar results.

Lastly, similar results can be observed when using the distillation-plus-pruning methods. From the beginning, DPWavLM can preserve much of the robustness of WavLM. However, if trained with RobustDistiller, we can further improve its performance in noisy settings and also its generalizability in clean scenarios, as observed for content- and speaker-related tasks.

\subsection{Effect of noise type in low SNR regimes}
\begin{figure}
   \centering
   \includegraphics[width=\linewidth]{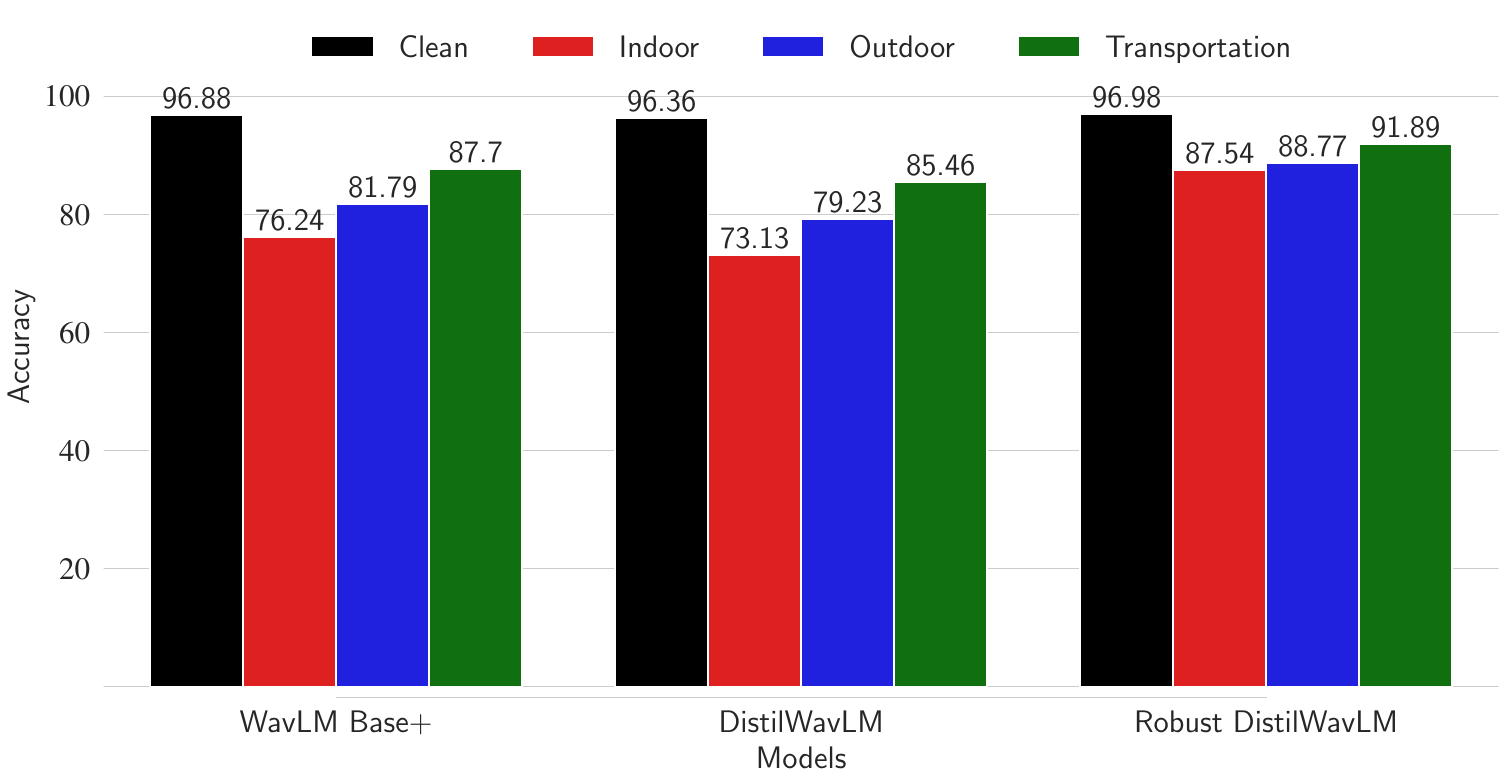}
   \caption{KS accuracy per noise type with clean speech levels used for referencing.}
   \label{fig:noise_env} \vspace{-3mm}
\end{figure}

Results above have shown the sensitivity of different upstream models to the presence of noise during testing. An in-depth analysis showed that this effect was particularly true at SNR regimes below an SNR of 10 dB. Here, we explore if noise type under these SNR regimes play a role on final system accuracy. For simplicity, here we focus only on the KS task and noise levels at the SNR range of $[-5,5]$ dB.  The DCASE2020 noise dataset used for testing was comprised of noise signals under three categories: indoor (e.g., shopping mall, airport), outdoor (e.g., park, public square), and transportation (e.g., bus, metro, tram). Figure~\ref{fig:noise_env} shows the accuracy achieved under the clean condition, as well as the three different noise types for the WavLM Base+ Teacher model, as well as this model distilled with the benchmark distiller and the proposed RobustDistiller frameworks. As can be seen, for the benchmark systems, indoor noise causes the most significant decline in accuracy, which could be attributed to the combined impact of reverberation and additive noise. Transportation noise had the least impact on the benchmark methods. The proposed approach, in turn, resulted not only in the highest overall accuracy across the three noise types, but showed to be independent of noise type, thus further validating the noise robustness properties of the proposed solution.

\subsection{Effects of room size}
\begin{figure}
   \centering
   \includegraphics[width=\linewidth]{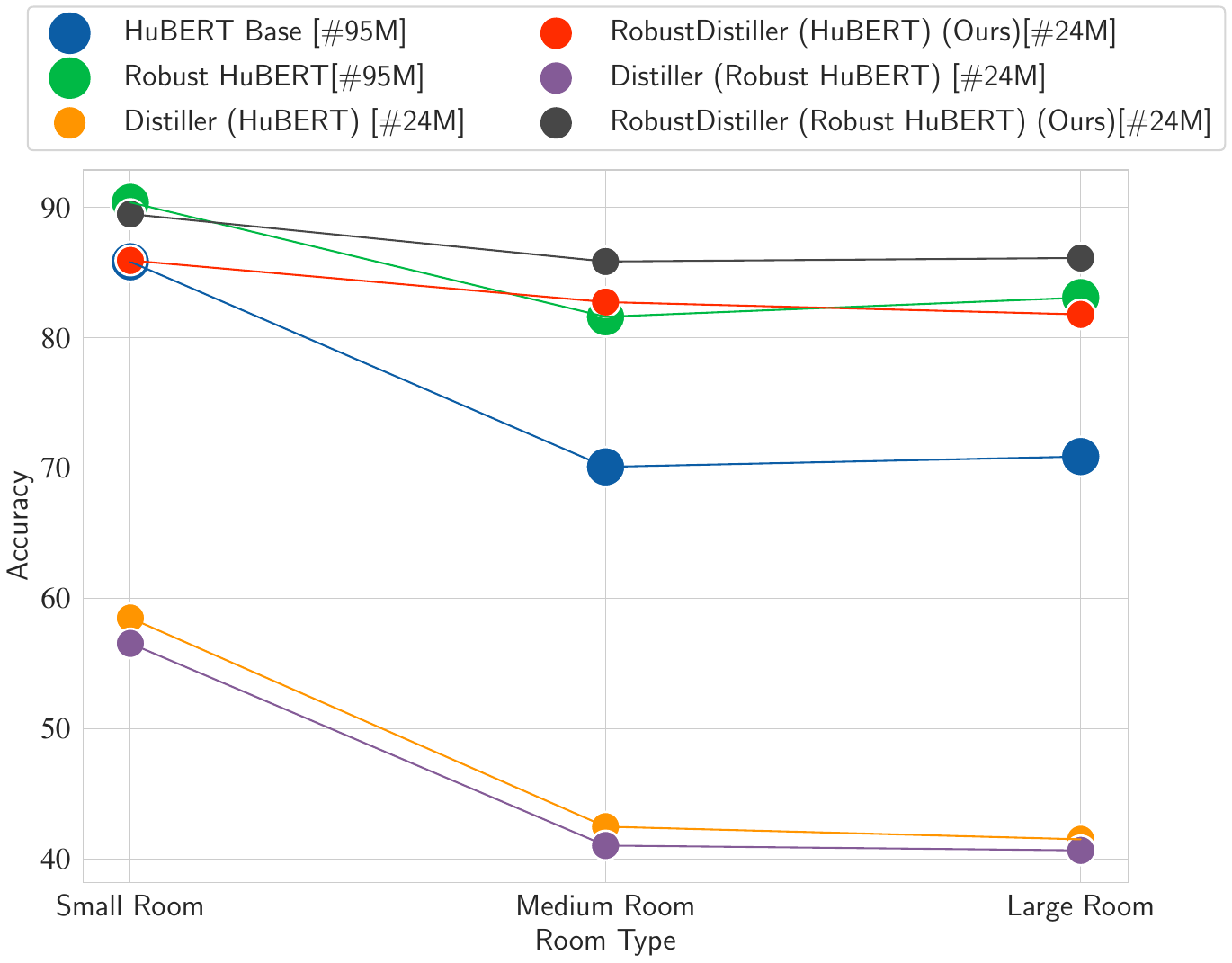}
   \caption{IC performance per room size for different HuBERT distillation schemes.}
   \label{fig:reveb_env} \vspace{-3mm}
\end{figure}

Lastly, we explore the impact of room size on system performance. Room sizes are directly proportional to reverberation times, with larger rooms typically resulting in lower performance. For this investigation, we use the IC task and HuBERT-based variations and the recorded RIRs from small, medium, and large-sized rooms, as described in Section~\ref{sec:data}. Figure~\ref{fig:reveb_env} depicts the accuracy achieved for each of the three room sizes. As expected, accuracy decreases as room sizes increase. The two proposed RobustDistiller setups achieved the highest accuracy across the three room sizes, with accuracy values at times outperforming the Teacher model, whilst requiring just one-quarter of the parameters. These findings are promising and suggesting that robust distillation for edge speech applications may indeed be possible via a robust distillation process.

\section{Conclusion}
In this work, we describe RobustDistiller, a methodology to improve the environmental robustness of the distillation process of speech representation learning algorithms. In particular, two modifications are proposed to original distillation recipes: (i) feature denoising knowledge distillation; and (ii) multi-task learning, where a speech enhancement step is jointly performed to reinforce the disentanglement of speech features from noise factors. We show that our recipe can be adapted to both layerwise distillation (e.g., DistilHuBERT) and distillation-plus-pruning (e.g., DPHuBERT) methods. We test the robustness of our model across clean and noisy conditions on top of 12 speech-processing tasks from SUPERB and SUPERB-SG. Experimental results show the proposed method frequently outperforming the original Teacher models with roughly four times fewer parameters. Overall, the proposed RobustDistiller on top of the DPWavLM recipe resulted in the best performance across tasks, thus suggesting that their use could be suitable for ``in-the-wild'' edge speech applications.




%

\bibliographystyle{IEEEtran}
\bibliography{refs}

\begin{thebibliography}{10}
\providecommand{\url}[1]{#1}
\csname url@samestyle\endcsname
\providecommand{\newblock}{\relax}
\providecommand{\bibinfo}[2]{#2}
\providecommand{\BIBentrySTDinterwordspacing}{\spaceskip=0pt\relax}
\providecommand{\BIBentryALTinterwordstretchfactor}{4}
\providecommand{\BIBentryALTinterwordspacing}{\spaceskip=\fontdimen2\font plus
\BIBentryALTinterwordstretchfactor\fontdimen3\font minus \fontdimen4\font\relax}
\providecommand{\BIBforeignlanguage}[2]{{%
\expandafter\ifx\csname l@#1\endcsname\relax
\typeout{** WARNING: IEEEtran.bst: No hyphenation pattern has been}%
\typeout{** loaded for the language `#1'. Using the pattern for}%
\typeout{** the default language instead.}%
\else
\language=\csname l@#1\endcsname
\fi
#2}}
\providecommand{\BIBdecl}{\relax}
\BIBdecl

\bibitem{9893562}
A.~Mohamed, H.-y. Lee, L.~Borgholt, J.~D. Havtorn, J.~Edin, C.~Igel, K.~Kirchhoff, S.-W. Li, K.~Livescu, L.~Maaløe, T.~N. Sainath, and S.~Watanabe, ``Self-supervised speech representation learning: A review,'' \emph{IEEE Journal of Selected Topics in Signal Processing}, vol.~16, no.~6, pp. 1179--1210, 2022.

\bibitem{baevski2020wav2vec}
A.~Baevski, Y.~Zhou, A.~Mohamed, and M.~Auli, ``wav2vec 2.0: A framework for self-supervised learning of speech representations,'' \emph{Advances in Neural Information Processing Systems}, vol.~33, pp. 12\,449--12\,460, 2020.

\bibitem{hsu2021hubert}
W.-N. Hsu, B.~Bolte, Y.-H.~H. Tsai, K.~Lakhotia, R.~Salakhutdinov, and A.~Mohamed, ``Hu{BERT}: Self-supervised speech representation learning by masked prediction of hidden units,'' \emph{IEEE/ACM Transactions on Audio, Speech, and Language Processing}, vol.~29, pp. 3451--3460, 2021.

\bibitem{chen2022wavlm}
S.~Chen, C.~Wang, Z.~Chen, Y.~Wu, S.~Liu, Z.~Chen, J.~Li, N.~Kanda, T.~Yoshioka, X.~Xiao \emph{et~al.}, ``Wavlm: Large-scale self-supervised pre-training for full stack speech processing,'' \emph{IEEE Journal of Selected Topics in Signal Processing}, 2022.

\bibitem{huang2022improving}
K.~P. Huang, Y.-K. Fu, Y.~Zhang, and H.~yi~Lee, ``{Improving Distortion Robustness of Self-supervised Speech Processing Tasks with Domain Adaptation},'' in \emph{Proc. Interspeech 2022}, 2022, pp. 2193--2197.

\bibitem{babu2021xls}
A.~Babu, C.~Wang, A.~Tjandra, K.~Lakhotia, Q.~Xu, N.~Goyal, K.~Singh, P.~von Platen, Y.~Saraf, J.~Pino \emph{et~al.}, ``Xls-r: Self-supervised cross-lingual speech representation learning at scale,'' \emph{arXiv preprint arXiv:2111.09296}, 2021.

\bibitem{chang2022distilhubert}
H.-J. Chang, S.-w. Yang, and H.-y. Lee, ``Distilhubert: Speech representation learning by layer-wise distillation of hidden-unit bert,'' in \emph{ICASSP 2022-2022 IEEE International Conference on Acoustics, Speech and Signal Processing (ICASSP)}.\hskip 1em plus 0.5em minus 0.4em\relax IEEE, 2022, pp. 7087--7091.

\bibitem{guimarães2023exploration}
H.~Guimarães, A.~Pimentel, A.~Avila, M.~Rezagholizadeh, and T.~H. Falk, ``An exploration into the performance of unsupervised cross-task speech representations for "in the wild'' edge applications,'' 2023.

\bibitem{peng23c_interspeech}
Y.~Peng, Y.~Sudo, S.~Muhammad, and S.~Watanabe, ``{DPHuBERT: Joint Distillation and Pruning of Self-Supervised Speech Models},'' in \emph{Proc. INTERSPEECH 2023}, 2023, pp. 62--66.

\bibitem{robustdistiller}
H.~R. Guimarães, A.~Pimentel, A.~R. Avila, M.~Rezagholizadeh, B.~Chen, and T.~H. Falk, ``Robustdistiller: Compressing universal speech representations for enhanced environment robustness,'' in \emph{ICASSP 2023 - 2023 IEEE International Conference on Acoustics, Speech and Signal Processing (ICASSP)}, 2023, pp. 1--5.

\bibitem{jang2017categorical}
\BIBentryALTinterwordspacing
E.~Jang, S.~Gu, and B.~Poole, ``Categorical reparameterization with gumbel-softmax,'' in \emph{International Conference on Learning Representations}, 2017. [Online]. Available: \url{https://openreview.net/forum?id=rkE3y85ee}
\BIBentrySTDinterwordspacing

\bibitem{yang2021superb}
S.-w. Yang, P.-H. Chi \emph{et~al.}, ``Superb: Speech processing universal performance benchmark,'' \emph{arXiv preprint arXiv:2105.01051}, 2021.

\bibitem{tsai-etal-2022-superb}
\BIBentryALTinterwordspacing
H.-S. Tsai, H.-J. Chang, W.-C. Huang, Z.~Huang, K.~Lakhotia, S.-w. Yang, S.~Dong, A.~Liu, C.-I. Lai, J.~Shi, X.~Chang, P.~Hall, H.-J. Chen, S.-W. Li, S.~Watanabe, A.~Mohamed, and H.-y. Lee, ``{SUPERB}-{SG}: Enhanced speech processing universal {PER}formance benchmark for semantic and generative capabilities,'' in \emph{Proceedings of the 60th Annual Meeting of the Association for Computational Linguistics (Volume 1: Long Papers)}, S.~Muresan, P.~Nakov, and A.~Villavicencio, Eds.\hskip 1em plus 0.5em minus 0.4em\relax Dublin, Ireland: Association for Computational Linguistics, May 2022, pp. 8479--8492. [Online]. Available: \url{https://aclanthology.org/2022.acl-long.580}
\BIBentrySTDinterwordspacing

\bibitem{feng2023superb}
T.-h. Feng, A.~Dong, C.-F. Yeh, S.-w. Yang, T.-Q. Lin, J.~Shi, K.-W. Chang, Z.~Huang, H.~Wu, X.~Chang \emph{et~al.}, ``Superb@ slt 2022: Challenge on generalization and efficiency of self-supervised speech representation learning,'' in \emph{2022 IEEE Spoken Language Technology Workshop (SLT)}.\hskip 1em plus 0.5em minus 0.4em\relax IEEE, 2023, pp. 1096--1103.

\bibitem{panayotov2015librispeech}
V.~Panayotov, G.~Chen, D.~Povey, and S.~Khudanpur, ``Librispeech: an asr corpus based on public domain audio books,'' in \emph{2015 IEEE international conference on acoustics, speech and signal processing (ICASSP)}.\hskip 1em plus 0.5em minus 0.4em\relax IEEE, 2015, pp. 5206--5210.

\bibitem{kahn2020libri}
J.~Kahn, M.~Rivi{\`e}re \emph{et~al.}, ``Libri-light: A benchmark for {ASR} with limited or no supervision,'' in \emph{Proc. IEEE ICASSP}, 2020, pp. 7669--7673.

\bibitem{hsu2021robust}
W.-N. Hsu, A.~Sriram, A.~Baevski, T.~Likhomanenko, Q.~Xu, V.~Pratap, J.~Kahn, A.~Lee, R.~Collobert, G.~Synnaeve \emph{et~al.}, ``Robust wav2vec 2.0: Analyzing domain shift in self-supervised pre-training,'' \emph{arXiv preprint arXiv:2104.01027}, 2021.

\bibitem{JMLR:v17:15-239}
\BIBentryALTinterwordspacing
Y.~Ganin, E.~Ustinova, H.~Ajakan, P.~Germain, H.~Larochelle, F.~Laviolette, M.~March, and V.~Lempitsky, ``Domain-adversarial training of neural networks,'' \emph{Journal of Machine Learning Research}, vol.~17, no.~59, pp. 1--35, 2016. [Online]. Available: \url{http://jmlr.org/papers/v17/15-239.html}
\BIBentrySTDinterwordspacing

\bibitem{wang2022lighthubert}
R.~Wang, Q.~Bai, J.~Ao, L.~Zhou, Z.~Xiong, Z.~Wei, Y.~Zhang, T.~Ko, and H.~Li, ``Lighthubert: Lightweight and configurable speech representation learning with once-for-all hidden-unit bert,'' \emph{arXiv preprint arXiv:2203.15610}, 2022.

\bibitem{lee22p_interspeech}
Y.~Lee, K.~Jang, J.~Goo, Y.~Jung, and H.~R. Kim, ``{FitHuBERT: Going Thinner and Deeper for Knowledge Distillation of Speech Self-Supervised Models},'' in \emph{Proc. Interspeech 2022}, 2022, pp. 3588--3592.

\bibitem{ashihara22_interspeech}
T.~Ashihara, T.~Moriya, K.~Matsuura, and T.~Tanaka, ``{Deep versus Wide: An Analysis of Student Architectures for Task-Agnostic Knowledge Distillation of Self-Supervised Speech Models},'' in \emph{Proc. Interspeech 2022}, 2022, pp. 411--415.

\bibitem{10022474}
K.-P. Huang, Y.-K. Fu, T.-Y. Hsu, F.~R. Gutierrez, F.-L. Wang, L.-H. Tseng, Y.~Zhang, and H.-y. Lee, ``Improving generalizability of distilled self-supervised speech processing models under distorted settings,'' in \emph{2022 IEEE Spoken Language Technology Workshop (SLT)}, 2023, pp. 1112--1119.

\bibitem{grill2020bootstrap}
J.-B. Grill, F.~Strub, F.~Altch{\'e}, C.~Tallec, P.~Richemond, E.~Buchatskaya, C.~Doersch, B.~Avila~Pires, Z.~Guo, M.~Gheshlaghi~Azar \emph{et~al.}, ``Bootstrap your own latent-a new approach to self-supervised learning,'' \emph{Advances in neural information processing systems}, vol.~33, pp. 21\,271--21\,284, 2020.

\bibitem{6639100}
M.~L. Seltzer, D.~Yu, and Y.~Wang, ``An investigation of deep neural networks for noise robust speech recognition,'' in \emph{2013 IEEE International Conference on Acoustics, Speech and Signal Processing}, 2013, pp. 7398--7402.

\bibitem{wang2022improving}
H.~Wang, Y.~Qian, X.~Wang, Y.~Wang, C.~Wang, S.~Liu, T.~Yoshioka, J.~Li, and D.~Wang, ``Improving noise robustness of contrastive speech representation learning with speech reconstruction,'' in \emph{ICASSP 2022-2022 IEEE International Conference on Acoustics, Speech and Signal Processing (ICASSP)}.\hskip 1em plus 0.5em minus 0.4em\relax IEEE, 2022, pp. 6062--6066.

\bibitem{941023}
A.~Rix, J.~Beerends, M.~Hollier, and A.~Hekstra, ``Perceptual evaluation of speech quality (pesq)-a new method for speech quality assessment of telephone networks and codecs,'' in \emph{2001 IEEE International Conference on Acoustics, Speech, and Signal Processing. Proceedings (Cat. No.01CH37221)}, vol.~2, 2001, pp. 749--752 vol.2.

\bibitem{musan2015}
D.~Snyder, G.~Chen, and D.~Povey, ``{MUSAN}: {A} {M}usic, {S}peech, and {N}oise {C}orpus,'' 2015.

\bibitem{Salamon:UrbanSound:ACMMM:14}
J.~Salamon, C.~Jacoby, and J.~P. Bello, ``A dataset and taxonomy for urban sound research,'' in \emph{22nd {ACM} International Conference on Multimedia (ACM-MM'14)}, Orlando, FL, USA, Nov. 2014, pp. 1041--1044.

\bibitem{ko2017study}
T.~Ko, V.~Peddinti, D.~Povey, M.~L. Seltzer, and S.~Khudanpur, ``A study on data augmentation of reverberant speech for robust speech recognition,'' in \emph{Proc. IEEE ICASSP}, 2017, pp. 5220--5224.

\bibitem{Mesaros2018_DCASE}
\BIBentryALTinterwordspacing
A.~Mesaros, T.~Heittola, and T.~Virtanen, ``A multi-device dataset for urban acoustic scene classification,'' in \emph{Proceedings of the Detection and Classification of Acoustic Scenes and Events 2018 Workshop (DCASE2018)}, November 2018, pp. 9--13. [Online]. Available: \url{https://dcase.community/documents/workshop2018/proceedings/DCASE2018Workshop\_Mesaros\_8.pdf}
\BIBentrySTDinterwordspacing

\bibitem{app132312571}
\BIBentryALTinterwordspacing
A.~Pimentel, H.~R. Guimarães, A.~Avila, and T.~H. Falk, ``Environment-aware knowledge distillation for improved resource-constrained edge speech recognition,'' \emph{Applied Sciences}, vol.~13, no.~23, 2023. [Online]. Available: \url{https://www.mdpi.com/2076-3417/13/23/12571}
\BIBentrySTDinterwordspacing

\bibitem{5713237}
C.~H. Taal, R.~C. Hendriks, R.~Heusdens, and J.~Jensen, ``An algorithm for intelligibility prediction of time–frequency weighted noisy speech,'' \emph{IEEE Transactions on Audio, Speech, and Language Processing}, vol.~19, no.~7, pp. 2125--2136, 2011.

\bibitem{le2019sdr}
J.~Le~Roux, S.~Wisdom, H.~Erdogan, and J.~R. Hershey, ``Sdr--half-baked or well done?'' in \emph{ICASSP 2019-2019 IEEE International Conference on Acoustics, Speech and Signal Processing (ICASSP)}.\hskip 1em plus 0.5em minus 0.4em\relax IEEE, 2019, pp. 626--630.

\bibitem{4967982}
T.~H. Falk and W.-Y. Chan, ``Modulation spectral features for robust far-field speaker identification,'' \emph{IEEE Transactions on Audio, Speech, and Language Processing}, vol.~18, no.~1, pp. 90--100, 2010.

\bibitem{Nagrani17}
A.~Nagrani, J.~S. Chung, and A.~Zisserman, ``Voxceleb: a large-scale speaker identification dataset,'' in \emph{INTERSPEECH}, 2017.

\end{thebibliography}


 

\begin{IEEEbiography}[{\includegraphics[width=1in,height=1.25in,clip,keepaspectratio]{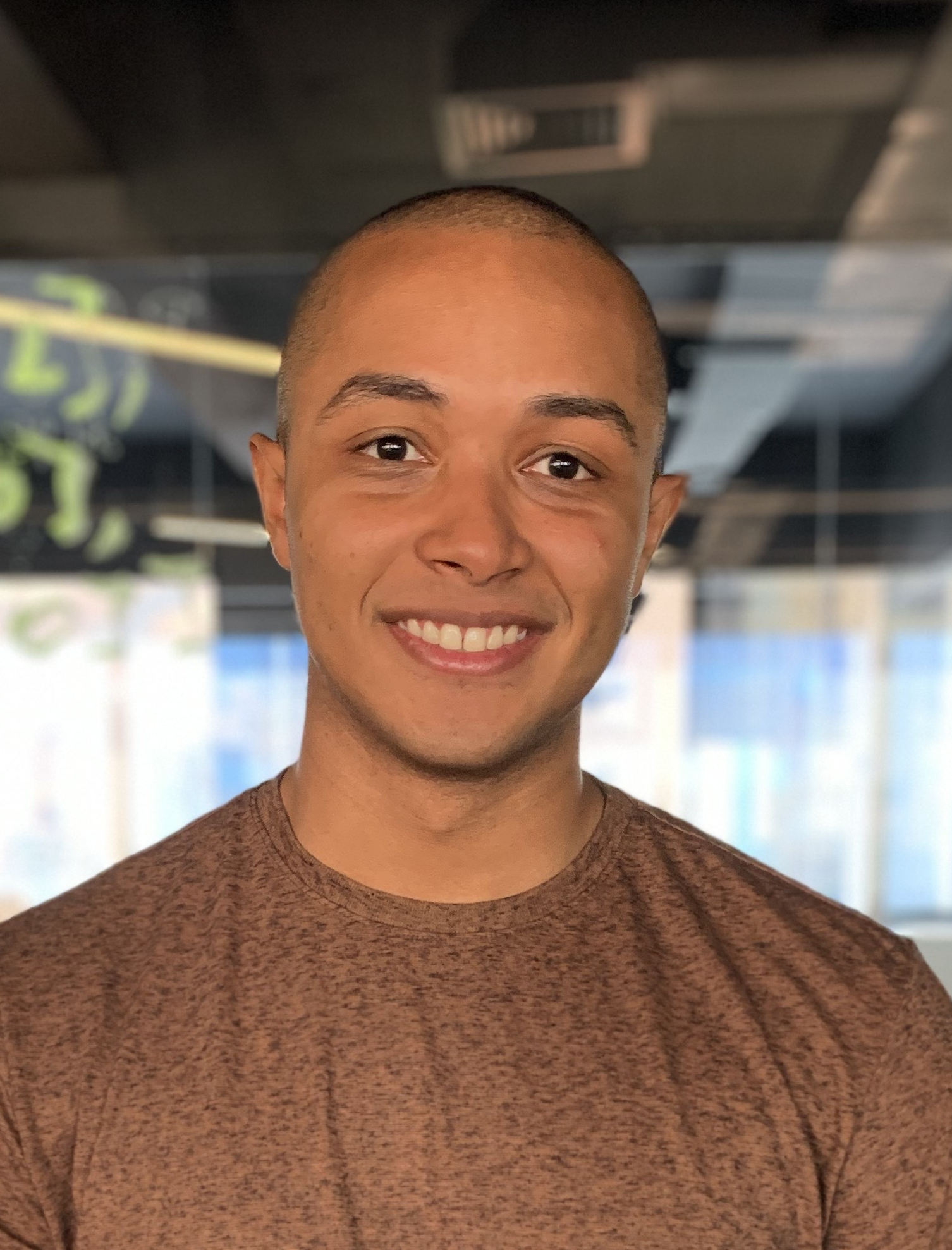}}]{Heitor R. Guimarães} (Student Member, IEEE) received his B.Eng. degree in Computer and Information Engineering from the Federal University of Rio de Janeiro (UFRJ); and the M.Sc. degree in Electrical Engineering from the University of São Paulo (USP). He is pursuing a Ph.D. in Telecommunications at the Institut National de la Recherche Scientifique (INRS - EMT). His research interests include self-supervised learning, domain adaptation, adversarial attacks, and efficient deep learning (e.g., model compression) for speech applications.
\end{IEEEbiography}

\begin{IEEEbiography}[{\includegraphics[width=1in,height=1.25in,clip,keepaspectratio]{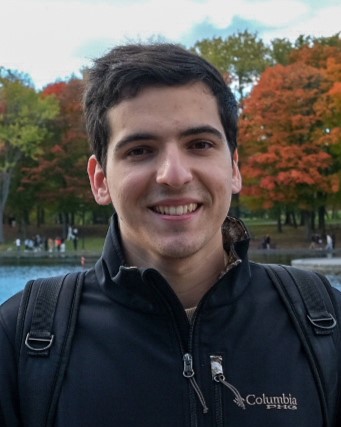}}]{Arthur Pimentel} received his B.Sc. degree in electronics engineering from the Federal University of Pernambuco (UFPE), Recife, Brazil, in 2021; and the M.Sc. degree in Telecommunications at the Institut National de la Recherche Scientifique (INRS - EMT). He is pursuing a Ph.D. in Telecommunications at INRS - EMT. His research interests include signal processing, self-supervised speech representation learning and deepfake detection. 
\end{IEEEbiography}

\begin{IEEEbiography}[{\includegraphics[width=1in,height=1.25in,clip,keepaspectratio]{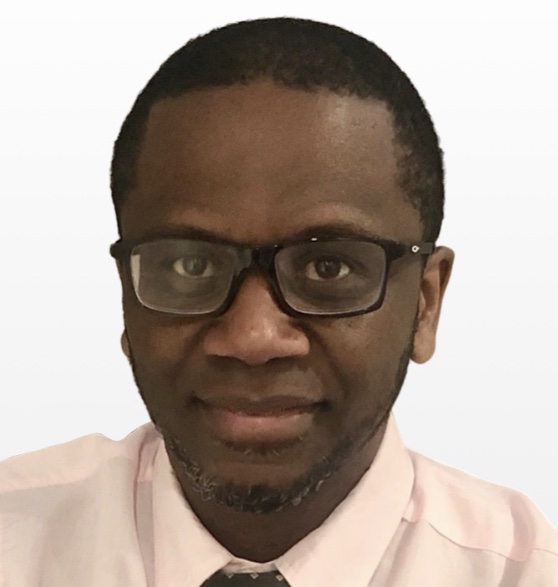}}]{Anderson Avila} is an Assistant Professor at INRS-EMT, working in the INRS-UQO Joint Research Unit in Cybersecurity. His research background is on machine learning and signal processing applied to natural language processing. During his PhD, Dr. Avila worked on the development of new models for speech quality assessment and on the robustness of voice biometrics. Prior to joining INRS-UQO, Dr. Avila was a researcher scientist in natural language and speech processing, working on projects related to model compression, low-latency and robustness of spoken language understanding. Dr. Avila received his BSc in Computer Science from the Federal University of São Carlos, his MSc from Federal University of ABC and his PhD from INRS.
\end{IEEEbiography}

\begin{IEEEbiography}[{\includegraphics[width=1in,height=1.25in,clip,keepaspectratio]{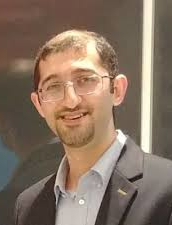}}]{Mehdi Rezagholizadeh} obtained a PhD from McGill University in Electrical and Computer Engineering (Centre of Intelligent Machines) in 2016. He joined Huawei in January 2017 and his research focus has been on different deep learning and NLP projects such as generative adversarial networks, neural machine translation, adversarial neural text generation, and efficient NLP for pre-trained language models. He is now a principal research scientist and he has been leading the NLP team of Huawei Noah's Ark Lab in Canada since 2018. Mehdi has contributed to more than 15 patents and 50 published papers in top journals and conferences. Moreover, Mehdi has organized several academic and industrial workshops such as NeurIPS 2021, 2022, and 2023 ENLSP workshops. He has served as technical committee member of ACL Rolling Review 21, ACL (22,23), EMNLP (20,21,22), NAACL 21, and EACL 21. 
\end{IEEEbiography}

\begin{IEEEbiography}[{\includegraphics[width=1in,height=1.25in,clip,keepaspectratio]{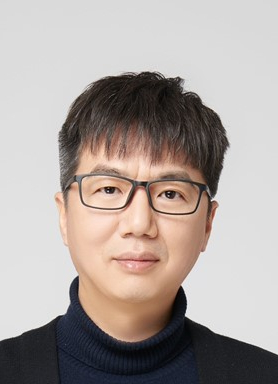}}]{Boxing Chen}  is the Chief Researcher of Montreal Research Centre for Huawai Technologies Canada, where he focuses on natural language processing and machine learning. Prior to joining Huawai, Dr. Chen held various positions, including Senior Staff Algorithm Expert at Alibaba Group's Machine Intelligence Lab and Research Officer at the National Research Council Canada (NRC), etc. He has co-authored over 100 papers in AI conferences and journals, received the "Best Paper Award" at MT Summit 2013 and a nomination for the same award at ACL 2013. He has also served as an area chair for conferences such as ACL, EMNLP, COLING, and AACL, and his teams have secured first place in over 20 machine translation competitions.
\end{IEEEbiography}

\begin{IEEEbiography}[{\includegraphics[width=1in,height=1.25in,clip,keepaspectratio]{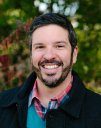}}]{Tiago H. Falk} received the B.Sc. degree in electronics engineering from UFPE, Recife, Brazil, in 2002, and the M.Sc. and Ph.D. degrees in electrical and computer engineering from Queen’s University, Kingston, ON, Canada, in 2005 and 2008, respectively. From 2008 to 2010, he was a Postdoctoral Fellow with the University of Toronto, Toronto, ON, Canada. Since 2010, he has been with INRS-EMT, University of Quebec, Quebec city, QC, Canada, where he is currently a tenured Full Professor, the Director of the Multimodal/Multisensory Signal Analysis and Enhancement Lab, and co-Director of the INRS-UQO Joint Research Unit on Cybersecurity. His research interests include signal processing and machine learning for next-generation secure human-machine interfaces.
\end{IEEEbiography}

\newpage

\vspace{11pt}

\vfill

\end{document}